\documentclass[%
 reprint,
 superscriptaddress,
 amsmath,amssymb,
]{revtex4-2}
\usepackage{graphicx}
\usepackage{dcolumn}
\usepackage{bm}
\usepackage{csquotes}
\usepackage[english]{babel}
\usepackage{siunitx}
\usepackage{booktabs}
\usepackage{tabularx}
\usepackage{adjustbox}
\usepackage{xcolor}

\usepackage{braket}
\usepackage[normalem]{ulem}
\usepackage[caption=false, justification=centerlast]{subfig}
\usepackage{todonotes}

\newcommand{\rom}[1]{\uppercase\expandafter{\romannumeral #1\relax}}

\usepackage[%
colorlinks=true,
urlcolor=blue,
linkcolor=blue,
citecolor=blue
]{hyperref}

\usepackage{tabularx}

\newcommand{\UOL}{Department of Physics, The University of Liverpool, Liverpool, L69 3BX, United Kingdom}
\newcommand{\CI}{Cockcroft Institute, Warrington WA4 4AD, United Kingdom}

\newcommand{\UOM}{Department of Physics and Astronomy, The University of Manchester, Manchester M13 9PL, United Kingdom}

\newcommand{\HIJ}{Helmholtz Institute Jena, Fr\"{o}belstieg 3, 07743 Jena, Germany}

\newcommand{\FSU}{Faculty of Physics and Astronomy, Friedrich-Schiller-Universit\"{a}t Jena, 07743 Jena, Germany}
\newcommand{\PKU}{Center for Applied Physics and Technology, HEDPS, and SKLNPT, School of Physics, Peking University, Beijing 100871, China}

\begin{document}
	      

\title{Theory of Relativistic Surface Plasmon Excitation on Smooth  Surface by High-Intensity Laser  
}

\author{Bifeng Lei}
\email{bifeng.lei@liverpool.ac.uk}
\affiliation{\UOL}
\affiliation{\CI}
%
%
%
%
%
%

\author{Bin Qiao}
\affiliation{\PKU}

\author{Matt Zepf}%
\affiliation{\HIJ}
\affiliation{\FSU}
\author{Guoxing Xia}
\affiliation{\UOM}
\affiliation{\CI}
\author{Carsten Welsh}
\affiliation{\UOL}
\affiliation{\CI}

\date{\today}

\begin{abstract}
We present a classical theory of relativistic surface plasmon (RSP) excitation at a smooth plasma-vacuum interface driven by either a ponderomotive force or an electric field of an intense laser pulse. 
Starting from Maxwell’s equations coupled to a cold-fluid plasma response, we derive a general driven wave equation for the RSP and solve it analytically. 
We show that an infinite planar surface enforces conservation of the in-plane wavevector. A finite longitudinal interaction length or axial modulation supplies a finite $k_z$ spectrum, while cylindrical curvature replaces one continuous transverse in-plane wavenumber by a discrete azimuthal mode index $m$. This partially relaxes the planar in-plane constraint, while axial phase matching remains controlled by the longitudinal spectrum of the drive.
The excitation strength is controlled by the overlap between the drive and the surface eigenfield, which is determined by the surface geometry. This provides a general principle for controlling RSP excitation. 
We also show that relativistic effects can substantially modify the dielectric response and can be preliminarily verified by particle-in-cell simulations. Within the local $\gamma_q$-modified dielectric model, the overlap-normalised planar source saturates at large $a_0$, and cylindrical curvature partially alleviates this reduction before strong surface softening develops.  
The role of surface geometry is analysed. A cylindrical surface can sustain an on-axis accelerating field, enabling highly nonlinear wakefield generation for particle acceleration. In addition, the cylindrical geometry imposes a precise mode-selection rule that provides intrinsic control over RSP excitation. 
Axisymmetric ponderomotive drive selects fundamental mode $m=0$. A linearly polarised laser field selects a superposition of $m=\pm 1$ modes, and a circularly polarised laser field selects a single helical mode. 
This theory can be extended to include other surface geometries and driving mechanisms and therefore presents a general framework. 
\end{abstract}

\maketitle

\section{Introduction}

When an electromagnetic (EM) wave propagates in a material medium, the field polarises the medium, thereby exciting a collective oscillation of free electrons. A quantum of such an oscillation is defined as a plasmon. 
A surface plasmon (SP) is a plasmon that occurs at the interface between a conductor (usually a metal) and a dielectric (such as vacuum or air). 
Originally, it is defined as an electron density wave rather than a propagating photon. 
When an SP couples with the incident EM wave and propagates along the surface, the resulting coupled mode is known as a surface plasmon polariton (SPP). SPP is a hybrid light–matter wave that contains both a photonic field and an electron plasma oscillation. 
Some authors use SP to denote the electrostatic limit of the SPP~\cite{Zayats:2005aa, agranovich2012surface}, whereas most use SP more broadly without making this distinction.

The study of SPs originated in non-relativistic contexts, primarily focusing on electrostatic surface modes in plasmas and solids. 
It was first observed in 1902 by R.W. Wood as an anomalous diffraction of light at a grating metal surface~\cite{Wood:1902aa}.  
In 1941, U. Fano provided the first theoretical investigation that established a connection between Wood's anomalies and the excitation of SPs~\cite{Fano:1941aa}.
The formal prediction of SP came from R.H. Ritchie in 1957, who invoked these collective electron oscillations to explain characteristic energy losses on thin metal surfaces~\cite{Ritchie:1957aa}. 
In 1960, E. A. Stern and R.A. Ferrell provided the first complete electrodynamic description of SPPs as propagating surface eigenmodes with a well-defined dispersion relation~\cite{Stern:1960aa}.
Late in 1965, A.A. Vedenov provided a general dispersion relation for surface plasma polaritons (SPPs) in a cold plasma half-space, showing frequencies ranging from $\omega_{\rm sp} \to \omega_p/\sqrt{2}$ down to $\omega_{\rm sp}=0$~\cite{Vedenov:1965aa}, where $\omega_p = \sqrt{4\pi n_e e^2 / m_e}$ is plasma frequency, $n_e$ electron density, $e$ electron charge and $m_e$ electron rest mass.
In 1968, A. Otto proposed that the problem of momentum mismatch between photons and plasmons can be solved by the development of prism-coupling techniques~\cite{Otto:1968aa}. This technique became central to the field of plasmonics~\cite{Yu:2019aa}.

The revolution of SP is driven by the advent of high-power lasers, which unlocked the relativistic regime of laser-matter interaction. The invention of Chirped Pulse Amplification (CPA) in 1985 by D. Strickland and G. Mourou~\cite{Strickland:1985aa} makes it possible to achieve high focal intensity (e.g. intensity higher than $10^{18}~\si{W/cm^2}$)~\cite{Mourou2019aa}, where the electron quiver motion becomes relativistic. As a result, the electron mass changes dramatically, and the magnetic component of the Lorentz force becomes significant. These relativistic effects have important consequences on the dynamics of surface electrons.  
Since 2000s, the experimental realisation of major relativistic laser-driven applications in high-order harmonic generation (HHG)~\cite{Carman:1981aa, Popov:1990aa, Zaidi:1991aa, Bulanov:1994aa,Lichters:1996aa, Gibbon:1996aa,Linde:1996aa, Melrose:2006aa, Baeva:2006aa}, ion acceleration~\cite{Snavely:2000aa, Umstadter:2003aa,Bigongiari:2011aa} and electron acceleration~\cite{ Willingale:2013aa,Riconda:2015aa, Fedeli:2016aa} has provided a strong motivation to understand and enhance the underlying energy coupling mechanisms with SPs in the relativistic regime, referred to as relativistic surface plasmons (RSPs).

With the increasing availability of high-power, high-contrast laser systems, recent experiments have demonstrated relativistic laser-grating surface interaction at intensities above $I > 10^{20}~\si{W/cm^{2}}$~\cite{Cerchez:2013aa, Ceccotti:2013aa, Jana:2024aa}. These developments have greatly stimulated both experimental~\cite{Willingale:2011aa,Raynaud:2018aa,Zhu:2020aa, Lad:2022aa} and theoretical~\cite{Hsu:2010aa, Brugge:2012aa, Raynaud:2018aa, Marini:2021aa} studies of RSP physics.
Nevertheless, conventional RSP schemes rely on coupling structures to externally compensate for momentum mismatch, and such structures are vulnerable to damage under high-power laser irradiation. 
Consequently, current research in this area is now advancing in two major directions. 
The first is the search for excitation mechanisms that emerge intrinsically from relativistic laser–plasma interactions, without the need for grating-assisted coupling. This is particularly important because intense laser pulses can damage structured surfaces, while pre-plasma formation further modifies the coupling conditions.
The second direction is the exploration of materials that can stably withstand high-power laser irradiation, since the existence of RSPs ultimately depends on the survival of the solid surface in an intense-field environment.

In 2021, X. Shen et al. proposed a new excitation mechanism of RSP via PIC simulations that does not require the traditional grating coupling ~\cite{Shen:2021aa}. This method is enabled by the scattering of a high-power laser pulse with the sharp end of a finite flat microtape. It also shows later that the high-flux X-ray radiation can be generated from the plasma-vacuum interface~\cite{Shen:2024aa}.
In 2022, J. Sarma et al. further demonstrated via simulations that efficient RSP electron acceleration using short pulse lasers currently available can occur in a flat foil irradiated at parallel or grazing incidence($\sim 5^o$ to the target surface) without surface modulation~\cite{Sarma:2022aa}.
In 2025, A. McCay et al. confirmed this mechanism experimentally, demonstrating that a 120 pC electron beam with a broad spectrum can be self-trapped in the RSP field and accelerated to 36 MeV along a flat, non-corrugated foil at parallel incidence~\cite{McCay:2025aa}. This is a milestone for smooth surface-bound RSP-based electron acceleration.

In contrast to these direct RSP-based acceleration schemes, B. Lei et al. proposed in the same year, through numerical simulations, the use of RSP-driven wakefields for electron acceleration~\cite{Lei:2025aa}. Rather than relying on a finite surface extent, their work showed that surface geometry itself can enable RSP excitation on a smooth surface. This method offers significant advantages. In particular, the RSP-based leaky and bubble wakefields supported by a cylindrical surface can provide both ultrahigh field strength and substantial structural flexibility. This is important as it enables both electron and positron beams to be accelerated in the electrostatic field of plasma wave driven by high-intensity lasers to achieve very high-quality beams.
They further proposed a beam-driven RSP wakefield acceleration scheme, in which the wakefield strength can reach up to $450~\si{TV/m}$ when driven by an ultra-relativistic, high-density electron beam~\cite{Lei:2025ab, Lei:2026intch}. At the same time, they suggested using structured nanomaterials, such as carbon nanotubes (CNTs), as a platform for sustaining extreme RSP excitation~\cite{Lei:2025aa, Bonatto:2023aa}. Owing to their excellent thermal and electronic properties, well-defined structure, large-area uniformity, high damage thresholds, and tailorable geometry, CNTs are particularly well suited for generating and studying RSPs in a highly controlled manner.
Also in the same year, B. Lei et al. theoretically and numerically demonstrated highly efficient RSP excitation for coherent synchrotron radiation generation, driven by a circularly polarised laser pulse on the inner surface of a near-critical-density microtube~\cite{Lei:2025oty}. They reported an acceleration gradient at the $\si{TeV/m}$-level. This points to a promising research direction, as it may offer a viable route toward the ultra-compact coherent radiation sources based on laser–solid interaction.

Despite these advances, the underlying physics of smooth-surface-bound RSPs and their associated wakefield excitation remains insufficiently understood, both in theoretical modelling and in numerical simulation. This lack of understanding creates a significant knowledge gap in the study of RSPs on general smooth surfaces and limits progress toward experimental validation and practical applications.

In this paper, we present a classical theory of RSP excitation driven by a finite laser pulse at a smooth plasma–vacuum interface.
We first develop the general theoretical framework from Maxwell’s equations and cold fluid equations. The wave equations for both the eigenmodes and their amplitudes are explicitly derived. 
We show that the surface geometry not only determines the form of the eigenfields, but also the real physical RSP field by coupling with the external drive. This provides a fundamental principle to manipulate the RSP excitation. On this basis, we further discuss the conditions for in-plane momentum conservation and for the breaking of translational symmetry.
The general theory is then applied to the planar and cylindrical surface with an external drive of ponderomotive or direct electric field of a Gaussian laser pulse. The explicit planar and axisymmetric cylindrical TM dispersion, eigenmode, and source terms are derived, while the non-axisymmetric cylindrical modes are treated through a TM-dominant approximation and exact azimuthal selection rules. Fully 3D particle-in-cell (PIC) simulations have been carried out to preliminarily verify theoretical results of the relativistic and curvature effects. 
The cylindrical surface effects are analytically studied, including dispersion shift and $m=0$ mode high-density cutoff. The curvature enables the mode selection, which provides great control for the RSP excitation and the surface electron modulation.
In principle, the theory presented in this paper can be directly extended to more general drive and surface geometries. As a result, it is capable of studying other RSP-related dynamics and effects in the field of classical plasma physics.

\section{Surface Plasmon in classical regime}
 Strictly speaking, SP should be considered in the context of quantum mechanics, where an SP is defined as a quantised excitation which involves a collective oscillation of surface charge and behaves like a particle with a discrete energy and momentum.
However, most important properties of SP can be satisfactorily described in a classical EM framework.
In a classical EM picture, an SP is defined as a fundamental EM mode supported at the interface between a material with negative permittivity and one with positive permittivity, and characterised by a well-defined resonance frequency~\cite{Bohren:2008aa}. 
Here, the dielectric function contains all the required information about the collective as well as the individual particle excitation. Since a boundary is inherently a macroscopic concept, its effects can also be naturally treated using classical electrodynamics. Therefore, the classical approach should be generally adequate if the objects supporting SPs are large enough compared to the mean free path of the conduction electrons.

The interpretation of surface plasmons (SPs) varies across different research fields. From the perspective of electrodynamics, an SP is an EM surface wave that exists because the metal exhibits a negative permittivity arising from electron plasma oscillations. In this sense, it can be regarded as a particular type of surface wave.
In optics, an SP is usually treated as an eigenmode of an interface.
In solid-state physics, an SP is viewed as a collective excitation of conduction electrons at an interface, with the material response described by an effective dielectric function rather than by an explicit microscopic model.
In plasma physics, the corresponding concept is often referred to as a surface plasma wave (SPW), namely a surface-bound EM mode supported by the interface between a plasma and another medium.
However, the term SPW alone is not sufficient to characterise the full physics considered in this work in a self-consistent way. 
First, our study involves crystalline metallic materials, such as metallic carbon nanotubes (CNTs), as the supporting medium. Such materials can sustain SPs through the response of conduction electrons even under relatively low-power laser irradiation, for example, in the MW range. This is precisely the regime in which the concept of SP was originally introduced.
Second, the laser–surface plasma interaction considered here involves not only electron density oscillations at the interface, but also their coupling to the external drivers.
For these reasons, we use the term RSP for SP excitation driven by relativistic laser pulses in this paper.

\section{Relativistic surface plasmon excitation on a solid plasma-vacuum surface}
 
\subsection{Basic equations and boundary conditions}

In the classical theory, the governing equations of RSP excitation come from Maxwell's theory with fluid response. 
We consider a solid plasma-vacuum interface of electron density $n_e(\bm{r})$, and the ions are immobile on the timescale of the RSP excitation. These assumptions are reasonable as a metal surface can be quickly ionised by the front of a high-intensity laser, and an ion is much heavier than an electron.
 The dynamics of RSP driven by an external drive $\bm{f}_{\rm ext}$ can be described by Maxwell's equations 
\begin{equation}
	\begin{split}
		& \nabla \cdot \bm{E} = 4 \pi \rho \mathrm{,} \quad \quad \nabla \times \bm{E} = -\frac{1}{c} \frac{\partial \bm{B}}{\partial t} \\
		&  \nabla \cdot \bm{B} = 0 \mathrm{,} \quad \quad \nabla \times \bm{B} = \frac{4\pi}{c} \bm{J} + \frac{1}{c} \frac{\partial \bm{E}}{\partial t} \mathrm{,}
	\end{split}
	\label{eq:Maxwell_eqs}
\end{equation}
and the relativistic cold-electron fluid momentum equation
\begin{equation}
	 m_e (\partial_t + \bm{v}\cdot \nabla) (\gamma \bm{v}) = -e \left( \bm{E} + \frac{\bm{v}}{c} \times \bm{B} \right) + \bm{f}_{\rm ext} \mathrm{,}
	\label{eq:momentum_equ}
\end{equation}
and the continuity equation
\begin{equation}
	\frac{\partial n_e}{\partial t} + \nabla \cdot (n_e \bm{v} )= 0 \mathrm{,}
\label{eq:continuity_equ}
\end{equation}
where $\bm{v}$ is the electron fluid velocity and $\gamma=(1-\bm{v}^2/c^2)^{-1/2}$. $\bm{E}$ and $\bm{B}$ are electric and magnetic fields. $\rho$ and $\bm{J}$ are charge and current density, respectively. $c$ is the speed of light in vacuum. All equations are written in CGS units.

The presence of matter modifies the EM fields that are described by two constitutive equations, for linear media, in the form $\bm{D} = \epsilon \bm{E}$, $\bm{B} = \mu \bm{H}$, where $\epsilon$ and $\mu$ are electric permittivity and magnetic permeability, respectively. For linear, isotropic and homogeneous materials, they are scalars. 
To obtain a full description of the EM field on the interface, the boundary conditions should be implemented as 
\begin{equation}
	\label{eq:boundary_cross}
	\begin{split}
		& \hat{n}_{12} \times (\bm{E}^{(2)} -\bm{E}^{(1)} ) = 0 \\
		& \hat{n}_{12} \times (\bm{H}^{(2)}- \bm{H}^{(1)}) = \frac{4 \pi}{c} \bm{J}_s \mathrm{,}
	\end{split}
\end{equation}
where $\hat{n}_{12}$ denotes a normal vector pointing from media 1 to media 2. The subscripts $i=1$ and $2$ refer to each of the bounding nonmagnetic media where $\mu_1=\mu_2=1$. 
Eq.~\eqref{eq:boundary_cross} states that the tangential component of $\bm{E}$ is continuous across this interface while the tangential component of $\bm{H}$ equals the surface current density $\bm{J}_s$ across the interface. 
Two scalar equations are
\begin{equation}
\label{eq:boundary_normal}
	\begin{split}
		& \hat{n}_{12} \cdot (\bm{D}^{(2)} - \bm{D}^{(1)}) = 4 \pi \rho_s \\
		& \hat{n}_{12} \cdot (\bm{B}^{(2)} - \bm{B}^{(1)}) = 0 \mathrm{,}
	\end{split}
\end{equation}
where $\rho_s$ denotes the surface charge. 
Although Eq.~\eqref{eq:boundary_cross} and \eqref{eq:boundary_normal} are written in their most general form, the eigenmode problem considered below contains no externally imposed sheet charge or sheet current. Therefore, when deriving the RSP dispersion relation, we set $\rho_s=0$ and $\bm{J}_s=0$ as independent external sources. The induced surface charge associated with the RSP is encoded in the discontinuity of the normal electric field through the dielectric response, while the normal displacement field satisfies the usual source-free boundary condition.

In strong laser fields, the electron quiver motion changes the effective inertia. A cycle-averaged quiver Lorentz factor $\gamma_q$ can be introduced to define a relativistically modified dielectric response in adiabatic approximation. 
For a laser with normalised strength  $a_s=e\bm{E}_L/m_e c \omega_L$, the quiver factor is 	$\gamma_q = \sqrt{1+s_{p} a_s^2}$, where $s_{p}=1$ for circular polarisation (CP) and $s_{p}=1/2$ for linear polarisation (LP). $\omega_L$ is the laser carrier frequency and $a_s$ is the representative laser amplitude at the surface or in the dominant RSP-overlap region.
For a weak laser pulse where $a_s \ll 1$, $\gamma_q\simeq 1 + s_{p} a_s^2/2$. 
For strong laser pulse where $a_s \gg 1$, $\gamma_q\sim  \sqrt{s_{p}} a_s$.
The relativistic quiver increases the effective electron mass as $m_{e}^{*}=\gamma_q m_e$ and reduces the plasma frequency $\omega^*_{p}=\omega_p/\sqrt{\gamma_q}$.

\subsection{Linear plasma response and surface currents}
In the low-energy regime, RSP is driven by surface charge.
The surface charge does not penetrate the metal as $\nabla \cdot \bm{P} =0$ below the interface, where $\bm{P}$ represents the polarisation of the medium. The contribution to the surface charge is a pure surface term. For a plasma-vacuum interface, it is simply given by the polarisation of the plasma $\bm{P}_{2}$. From a physical point of view, a surface charge must have some finite extension along the normal direction. One has to account for non-local effects to introduce the relevant length scale, namely the Thomas-Fermi length. It is of the order of tenths of $\si{nm}$, which is negligible on the scale that matters to us in the $\si{\mu m}$ scale~\cite{Greffet:2012aa}. 
 In the strong-field regime, the dynamics of the RSP are more naturally described in terms of the driven surface-current response. Within the present local bulk model, this current is carried where $n_0 \neq 0$. The vacuum-side dynamics would require an extended model including electron trapping and emission from the surface.

We consider a solid plasma-vacuum interface with equilibrium electron density $n_0(\bm{r})$. On the timescale of the RSP motion, the ions are taken to be immobile, so the equilibrium ion density satisfies $n_i(\bm{r})=n_{0}(\bm{r})$. The total electron density can be written as $n_e(\bm{r},t)=n_0(\bm{r}) + \delta n (\bm{r},t)$, where $|\delta n| \ll n_0$ in the linear-response stage.
The current can be separated into two parts as $\bm{J} = \bm{J}_{\rm{pol}} + \bm{J}_{\rm{ext}}$, where $\bm{J}_{\rm{pol}}$ represents the linear plasma response to drive and $\bm{J}_{\rm{ext}}$ the current corresponding to the external drive.
The linear plasma response is obtained by linearising Eq.~\eqref{eq:momentum_equ} and neglecting the magnetic and convective nonlinear terms. In the frequency domain $e^{ik_z z - i \omega t}$, we get
\begin{equation}
	i\omega m_e \gamma_q \bm{v}_{\rm{pol}} = e \bm{E}
\end{equation}
and 
\begin{equation}
	i\omega m_e \gamma_q\bm{v}_{\rm{ext}} =  - \bm{f}_{\rm ext}
\end{equation}
where the fluid velocity is given by $\bm{v}_{\rm{pol}} = e\bm{E}/(i \omega m_e \gamma_q)$. $\bm{v}_{\rm{ext}}$ denotes the velocity associated with the source current. $k_z$ is defined as the propagation constant of the surface mode along $z$ and depends on the surface geometry. $\omega$ is the Fourier frequency of the driving component under consideration. The vacuum wavenumber (magnitude of the wavevector in free space) for the frequency is given as $k_0=\omega /c$.
As a result, the linear current is 
\begin{equation}
	\bm{J}_{\rm pol} = -\frac{e^2 n_0 (\bm{r})}{i \omega m_e \gamma_q} \bm{E} = i \frac{\omega_p^2}{4\pi \omega \gamma_q} \bm{E} \mathrm{,}
	\label{eq:linear_current} 
\end{equation}
The external velocity is
\begin{equation}
\bm{v}_{\rm{ext}} = - \frac{1}{i \omega m_e \gamma_q} \bm{f}_{\rm ext} \mathrm{,}
\end{equation}
which corresponds to the source current as
\begin{equation}
	\bm{J}_{\rm{ext}} = -e n_0(\bm{r}) \bm{v}_{\rm{ext}} = -i \frac{e n_{0}(\bm{r})}{m_e \omega \gamma_q} \bm{f}_{\rm ext} \mathrm{.}
	\label{eq:ext_current}
\end{equation}

By identifying $\bm{J}_{\rm{pol}} = \partial \bm{P} / \partial t = - i \omega \bm{P}$, and the electric polarisation $\bm{P}=\left [\varepsilon(\omega)-1 \right ] \bm{E} / 4 \pi$, we get the relativistic Drude permittivity in plasma as
\begin{equation}
	\varepsilon_2(\omega) = 1 - \frac{\omega^2_p}{\gamma_q \omega (\omega+i \nu)} \rm{,}
	\label{eq:drude_perm}
\end{equation}
where $\nu$ represents the collision rate in plasma. In the eigenmode calculation, $\gamma_q$ is treated as a frozen cycle-averaged parameter evaluated at the relevant surface or overlap region.  
In what follows, $\gamma_q$ is held constant in $\varepsilon(\omega)$, mode quantities and in the source-current response. The prescribed drive still retains its explicit spatial-temporal envelope, such as the Gaussian factors used below.
It is used to define the instantaneous RSP eigenmode.
In vacuum, $\varepsilon_1 = 1$. Thus, at low frequency $\omega$, the plasma behaves as a dispersive medium for RSP excitation.
This formula in Eq.~\eqref{eq:drude_perm} contains all the material responses and is the entire bridge from fluid model to EM surface mode excitation, where RSP can be treated as an EM surface eigenmode of a medium with negative effective permittivity.
 
\subsection{General driven wave equation}
From the Maxwell-Amp\`{e}re equation, the Helmholtz equation for $\bm{E}$ is
\begin{equation}
		\left [ \nabla \times \nabla \times - k_0^2 \varepsilon(\bm{r}, \omega) \right ]\bm{E} = \mathcal{L}(\omega, k_z) \bm{E} =  i \frac{4 \pi \omega}{c^2} \bm{J}_{\rm{ext}}  \mathrm{,}
\label{eq:driven_wave_equ}
\end{equation}
where the operator is defined as $\mathcal{L} (\omega, k_z) \equiv \nabla \times \nabla \times - k_0^2 \varepsilon(\bm{r},\omega)$.
The source term is given by 
\begin{equation}
	\bm{S}_{\rm ext} = i \frac{4 \pi \omega}{c^2} \bm{J}_{\rm{ext}} = \frac{4 \pi e n_0(\bm{r}_{\perp})}{ m_e c^2 \gamma_q} \bm{f}_{\rm ext}\rm{.}
	\label{eq:source_general_01}
\end{equation} 
Eq.~\eqref{eq:driven_wave_equ} can be written in the simple form as 
\begin{equation}
	\mathcal{L}(\omega, k_z) \bm{E} = \bm{S}_{\rm ext} \rm{.}
	\label{eq:forced_wave_eq_full_1}
\end{equation}

\begin{figure}
	\centering
	\includegraphics[width=0.48\textwidth]{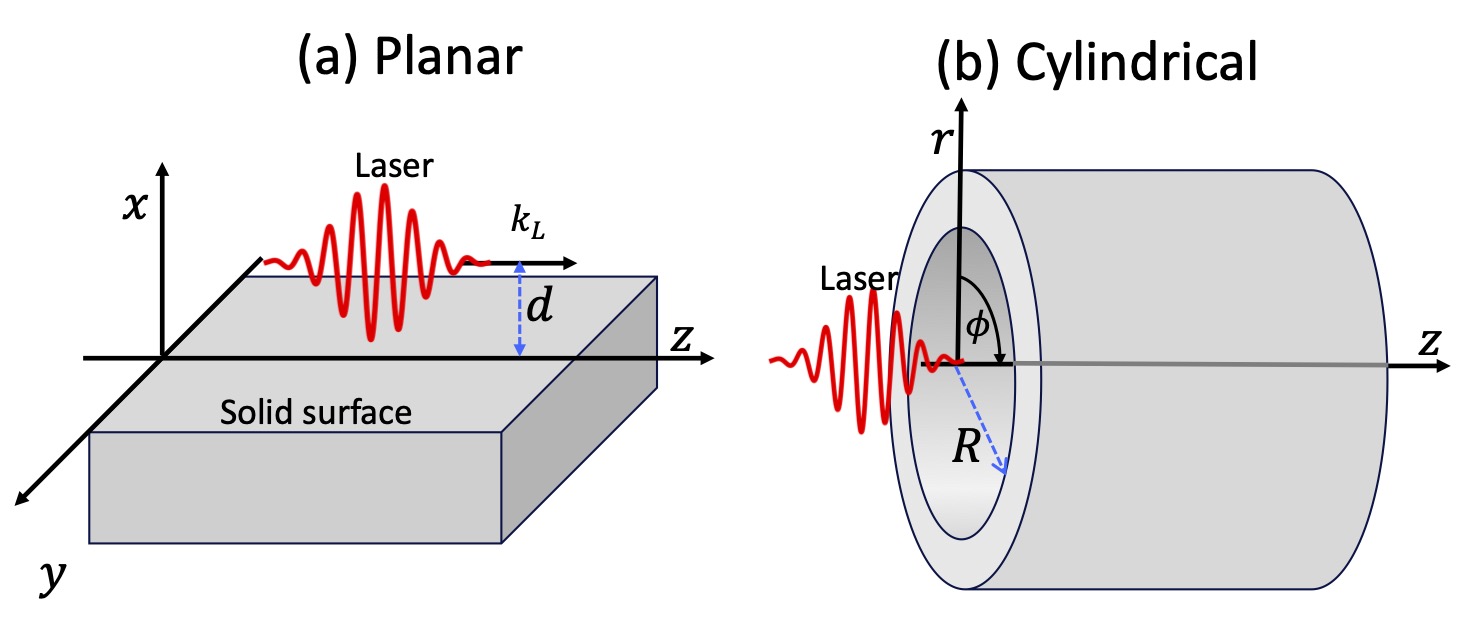}
	\caption{(a) Planar geometry. The interface lies at $x=0$ and the laser pulse propagates along $z$ at $x=d$.  (b) Cylindrical geometry. The interface lies at $r=R$ and the laser pulse propagates along axis, $r=0$.}
	\label{fig:geometry}
\end{figure}

Therefore, the bound surface eigenmodes $\{\bm{E}_{sp}\}$ satisfy 
\begin{equation}
	\mathcal{L}(\omega_{sp}, k_{sp}) \bm{E}_{sp} (\bm{r}_{\perp}) =0 \mathrm{,}
	\label{eq:eigenmodes_equs}
\end{equation}
with the propagation factor $e^{i\zeta_{sp}}$, where $\zeta_{sp}=k_{sp} z - \omega_{sp} t$. $\bm{r}_{\perp}$ denotes coordinates transverse to $z$. $\omega_{sp}$ and $k_{sp}$ are the eigenfrequency and propagation wavenumber of the eigenmode. Eq.~\eqref{eq:eigenmodes_equs} indicates that the solution of eigenmodes depends on the surface geometry. 
The physical field can be written as
\begin{equation}
	\bm{E}(\bm{r}, t) \approx \Re \left [ A(z,t) \bm{E}_{\rm{sp}}(\bm{r}_{\perp}) e^{i \zeta_{sp}} \right ] \rm{,}
	\label{eq:physical_E_field}
\end{equation}
where $A(z,t)$ denotes the envelope and is obtained by solving the full forced wave equation in Eq.~\eqref{eq:forced_wave_eq_full_1}.
Note that in the planar case, the full symmetry analysis is naturally two-dimensional along the interface, but in the case below, we restrict to propagation along $z$ as shown in Fig.~\ref{fig:geometry}, so that the RSP amplitude is written as $A(z,t)$.
For dispersive and lossless media $\nu=0$, the correct energy normalisation is the mode energy
\begin{equation}
	\mathcal{N}_{sp} = \frac{1}{16 \pi} \int \left( |\bm{B}_{sp}|^2 + \Re\{\partial_{\omega}(\omega \varepsilon) \} |\bm{E}_{sp}|^2 \right ) d \mathbb{A} \mathrm{,}
\end{equation}
where $\int d \mathbb{A}$ denotes the integration over the cross-section transverse to $z$ direction. For cylindrical geometry, $d \mathbb{A}=r dr d\phi$. For planar geometry with translational invariance in $y$, the normalisation is taken per unit length in $y$, so $d \mathbb{A}=dx$.

Assuming slowly varying envelope (SVEA), $|\partial_z A| \ll k_{sp} |A|$ and $|\partial_t A| \ll \omega_{sp} |A|$, we will only keep the first order in $\partial_z A$ and $\partial_t A$. 
In homogeneous eigen problem, $\partial_z \to i k_{sp}$ and $\partial_t \to - i\omega_{sp}$, 
\begin{equation}
	\begin{split}
		& \partial_t(A e^{i \zeta_{sp}}) = e^{i\zeta_{sp}} (-i \omega_{sp} A + \partial_t A) \\
		& \partial_z(A e^{i \zeta_{sp}}) = e^{i\zeta_{sp}} (i k_{sp}A + \partial_z A) \rm{,} 
	\end{split}
\end{equation}
therefore, the envelope introduces small shifts in $\omega_{sp}$ and $k_{sp}$ as $\omega_{sp} \to \omega_{sp} + i\partial_t$ and $k_{z} \to k_{sp} - i\partial_z$.
By expanding $\mathcal{L}(\omega, k_z)$ to first order about $(\omega_{sp}, k_{sp})$, we get
\begin{equation}
\begin{split}
		& e^{i \zeta_{sp}} \bigg \{ \mathcal{L}(\omega_{sp}, k_{sp}) [A \bm{E}_{sp}]  +  i\bigg [\partial_{\omega} \mathcal{L} \bigg]_{\rm sp}  \partial_t [A \bm{E}_{sp}] \\
		& - i\bigg [\partial_{k_z} \mathcal{L} \bigg]_{\rm sp} \partial_z [A \bm{E}_{\rm sp}] \bigg \} = i \frac{4 \pi \omega_{sp}}{c^2} \bm{J}_{\rm res} \rm{,}
		\label{eq:first_derivation_equ}
\end{split}
\end{equation}
where $\bm{J}_{\rm ext}$ can be written as $\bm{J}_{\rm ext}=\bm{J}_{\rm res} (\bm{r}_{\perp},z,t) e^{i\zeta_{sp}} + \bm{J}_{\rm non}$ with the non-resonant term $\bm{J}_{\rm non}$. 
The zeroth-order eigenmode term vanishes at the eigenpoint $\mathcal{L}(\omega_{sp}, k_{sp}) [A \bm{E}_{sp}]=0$.
Taking the scalar product of Eq.~\eqref{eq:first_derivation_equ} with $\bm{E}^{*}_{sp}$ and integrate over the transverse cross-section gives
\begin{equation}
\begin{split}
		i\partial_t A \int d \mathbb{A} \, \bm{E}^{*}_{\rm sp} \cdot &  [\partial_{\omega} \mathcal{L} ]_{\rm sp} \bm{E}_{\rm sp}  -  i\partial_z A  \int d \mathbb{A} \, \bm{E}^{*}_{\rm sp} \cdot [\partial_{k_z} \mathcal{L} ]_{\rm sp}  \bm{E}_{\rm sp}   \\
		& = i \frac{4 \pi \omega_{\rm sp}}{c^2} \int d \mathbb{A} \, \bm{E}^{*}_{sp} \cdot \bm{J}_{\rm res} \rm{.}
		\label{eq:mode_projection}
\end{split}
\end{equation}
We have the explicit form of $[\partial_{\omega}]_{\rm sp}$ as
\begin{equation}
	[\partial_{\omega} \mathcal{L}]_{\rm sp} = -\frac{1}{c^2} \bigg ( 2 \omega_{sp} \varepsilon + \omega_{sp}^2  \partial_{\omega} \varepsilon \bigg) \bm {I}\mathrm{,}
\end{equation}
where $\bm{I}$ represents the identity operator. 
For the reciprocal lossless eigenproblem used here, the adjoint mode equals the complex-conjugate mode, $\bm{E}_{sp}^{\dag}=\bm{E}_{sp}^{*}$. All surface terms arising from integration by parts vanish because the RSP mode is bound and satisfies the interface boundary conditions.
From the eigen-equations in Eq.~\eqref{eq:eigenmodes_equs} and vector calculus, we can get two reciprocity identities
\begin{equation}
	\int d \, \mathbb{A} \bm{E}_{\rm sp}^{*} \cdot [\partial_{\omega} \mathcal{L}]_{\rm sp} \bm{E}_{\rm sp} = - \frac{16 \pi \omega_{sp}}{c^2} \mathcal{N}_{sp} \rm{,}
	\label{eq:brillouin_iden}
\end{equation}
and the group-velocity identity
\begin{equation}
	\int d \, \mathbb{A} \bm{E}_{\rm sp}^{*} \cdot [\partial_{k_z} \mathcal{L}]_{sp} \bm{E}_{\rm sp} = \frac{16 \pi \omega_{sp}}{c^2} v_g \mathcal{N}_{sp} \rm{,}
	\label{eq:group_velocity_iden}
\end{equation}
where $v_g=(\partial \omega / \partial k_z)_{\rm sp}$ is the group velocity and then $v_g \mathcal{N}_{sp}$ is the energy flow.
With the convention in Eq.~\eqref{eq:physical_E_field}, these identities imply that the projected source enters with a prefactor $-1/(4\mathcal{N}_{sp})$.
By inserting Eq.~\eqref{eq:brillouin_iden} and \eqref{eq:group_velocity_iden} into Eq.~\eqref{eq:mode_projection}, the driven wave equation becomes
\begin{equation}
	(\partial_t + v_g \partial_z) A = S(z, t) \rm{.}
	\label{eq:driven_wave_equ_source}
\end{equation}
With the source term  
\begin{equation}
	S(z, t) = - \frac{1}{4 \mathcal{N}_{sp}} \int d \mathbb{A} \, \bm{E}_{\rm sp}^{*} (\bm{r}_{\perp}) \cdot \bm{J}_{\rm res}(\bm{r}_{\perp},z,t) \mathrm{,}
	\label{eq:source_term}
\end{equation}
and adding a small loss $\Gamma$, Eq.~\eqref{eq:driven_wave_equ_source} becomes
\begin{equation}
	(\partial_t + v_g \partial_z + \Gamma ) A(z,t) = - \frac{1}{4 \mathcal{N}_{sp}} \int d \mathbb{A} \, \bm{E}_{\rm sp}^{*} (\bm{r}_{\perp}) \cdot \bm{J}_{\rm res}(\bm{r}_{\perp},z,t)  \mathrm{,}
	\label{eq:imhomo_wave_equ_slowenv_full}
\end{equation}
which shows that the RSP envelope $A$ is transported at its group velocity $v_g$ and damped at rate $\Gamma$. 
$\Gamma$ is an effective modal damping rate and is not necessarily identical to the microscopic collision rate $\nu$ used in Eq.~\eqref{eq:drude_perm}.
Replacing $\bm{J}_{\rm ext}$ by the general force form in Eq.~\eqref{eq:ext_current} gives the general amplitude equation by
\begin{equation}
\begin{split}
	(\partial_t + v_g \partial_z + \Gamma ) A = 
	&  \frac{i e }{4 \omega_d \gamma_q m_e \mathcal{N}_{sp}}  
	\int d \mathbb{A} \, n_0(\bm{r}_{\perp})  \\
	& \bm{E}_{\rm sp}^{*} (\bm{r}_{\perp}) \cdot \bm{f}_{\rm res}(\bm{r}_{\perp},z, t)   \mathrm{,}
\end{split}
	\label{eq:imhomo_wave_equ_slowenv_full_2}
\end{equation}
where $\omega_{d}$ denotes the spectral frequency of the forcing component that couples to the mode. $f_{\rm res}$ denotes the component of the external force whose spatial and temporal phase matches the RSP eigenphase.
For example, for direct laser-field drive $\omega_d=\omega_L$. For cycle-averaged ponderomotive drive, the source is broadband in the slow-envelope spectrum, and after projection onto the SP mode, the resonant contribution is evaluated at $\omega_d=\omega_{sp}$.
Eq.~\eqref{eq:imhomo_wave_equ_slowenv_full_2} indicates that RSP excitation is controlled by the overlap between the eigenmodes and the external drive. 
On the other hand, the surface geometry can determine the efficiency of RSP excitation. 
It therefore provides a method of controlling RSP excitation through the surface geometry and the form of the driver, as discussed later. 

\subsection{In-plane momentum conservation and translational symmetry breaking}

Eq.~\eqref{eq:imhomo_wave_equ_slowenv_full_2} shows that the driven amplitude is determined by the Fourier component of the source that has the same phase as the surface eigenmode. In an infinite planar geometry, translational symmetry along the interface enforces conservation of the in-plane wavevector.
On a planar interface spanning over the $yz$-plane, because $\varepsilon$ depends only on $x$, the operator $\mathcal{L}(\omega, k_z)$ is invariant under any translation along the interface. Equivalently, the commutator $[\mathcal{L}(\omega, k_z), i \nabla_{\parallel}]=0$ where $\nabla_{\parallel}=\hat{y} \partial_y + \hat{z} \partial_z$. So the tangential wavevector is conserved.
The eigenmodes can be chosen as simultaneous eigenfunctions as $\bm{E}_{q_{\parallel}} \sim e^{i\bm{\bm{r}_{\parallel}} \cdot \bm{q}_{\parallel}}$ where $\bm{r}_{\parallel}=(y,z)$ denotes the coordinates parallel to the planar interface. Note that, to avoid ambiguity, we reserve $\bm{q}_{\parallel}$ for the general in-plane Fourier variable in planar geometry, while $k_z$ and $k_{sp}$ denote the axial propagation constant and the exact surface-mode eigenvalue, respectively.

Let's take one monochromatic incident plane wave in vacuum as $\bm{E}_i (\bm{r},t)=\bm{E}_{i,0}e^{i(k_x^{(i)}x + \bm{k}_{\parallel}^{(i)} \cdot \bm{r}_{\parallel} - \omega t)}$ where its tangential wavevector $|\bm{k}_{\parallel}^{(i)}|\leq \omega /c$ is a fixed constant.
From Eq.~\eqref{eq:ext_current}, the source current has the same tangential phase as $\bm{J}_{\rm ext} \propto e^{i \bm{k}_{\parallel}^{(i)} \cdot \bm{r}_{\parallel}}$. 
We can write the source term as
\begin{equation}
	S_{\bm{q}_{\parallel}} \propto \delta^{(2)} (\bm{q}_{\parallel} - \bm{k}_{\parallel}^{(i)}) \mathrm{,}
	\label{eq:general_selection_rule}
\end{equation}
where $\delta^{(2)}$ represents the 2D delta function.
Eq.~\eqref{eq:general_selection_rule} indicates that single plane wave can only drive the mode with the same in-plane wavevector. This is the selection rule of RSP excitation on an infinite planar interface. 
However, this constraint does not intrinsically come from Maxwell's equations. Instead, Maxwell's equations do admit the RSP as a homogeneous bound solution. It disappears once the translational symmetry is broken.
If the interface is not perfectly invariant along the surface, the source is no longer a single Fourier component~\cite{Stegeman:1983aa}. Instead, 
\begin{equation}
	\bm{J}_{\rm ext}(x,\bm{\bm{r}_{\parallel}}) = \int \frac{d^2 \bm{q}_{\parallel}}{(2\pi)^2} \tilde{\bm{J}}(x,\bm{q}_{\parallel}) e^{i\bm{q}_{\parallel} \cdot \bm{\bm{r}_{\parallel}}} \mathrm{.}
\end{equation}
Then the source becomes $S_{q_{\parallel}}\propto \tilde{\bm{J}}(x,\bm{q}_{\parallel})$ rather than a delta function at one fixed $\bm{k}_{\parallel}^{(i)}$. For example, a surface grating structure can add reciprocal lattice vectors $\bm{G}$, so $\bm{q}_{\parallel}=\bm{k}_{\parallel}^{i} + \bm{G}$. 
Curvature or finite size can also remove exact $\bm{q}_{\parallel}$ conservation. 
For example, a cylindrical geometry can replace the continuous planar $\bm{q}_{\parallel}$ by geometry-controlled surface eigenvalues, as discussed later.
In the later 1D envelope treatment, we restrict to propagation along $z$, so that $q_y=0$ and $q_{\parallel} \to k_z$.

\section{With ponderomotive drive }

For a general geometry, consider a laser pulse with a ponderomotive potential $U_p$ as $\bm{f}_{\rm ext}= - \nabla U_p$. 
Then, Eq.~\eqref{eq:imhomo_wave_equ_slowenv_full_2} becomes
\begin{equation}
\begin{split}
		(\partial_t + & v_g \partial_z + \Gamma ) A  = -\frac{ie}{4 \omega_{sp} \gamma_q m_e \mathcal{N}_{sp}} \int d \mathbb{A}  \\
		& \cdot n_0(\bm{r}_{\perp}) \bm{E}_{\rm sp}^{*}(\bm{r}_{\perp}) \cdot [\nabla U_p(\bm{r}_{\perp}, z, t)]_{\rm res}\mathrm{,}
\end{split}
	\label{eq:force_wave_equ_slowenv_full}
\end{equation}
where the resonant contribution to the projected amplitude equation is evaluated at $\omega_d=\omega_{sp}$. 
For a finite pulse, the resonant ponderomotive source is the Fourier component satisfying both the temporal and axial phase conditions $\Omega=\omega_{sp}$ and $q_z=k_sp$. 
For a source depending primarily on $\zeta_L=z-v_L t$, its spectrum satisfies approximately $\Omega=q_z v_L$, so efficient excitation requires small detuning $\Delta_p = \omega_{sp} - k_{sp} v_L$.
Since SPs are TM-like, the dominant contribution is usually the overlap with the normal electric field. In this approximation,
\begin{equation}
	\bm{E}_{\rm sp}^{*} \cdot \nabla U_p \approx \bm{E}_{\rm{sp},\perp}^{*} \cdot \nabla _{\perp} U_p \rm{.}
\end{equation}
The system thus can be determined by eigenmode $(\bm{E}_{sp}$, $\mathcal{N}_{sp}$, $v_g$) and the laser ponderomotive potential $U_p$.

In planar geometry, the coupling is controlled by the $\bm{q}_{\parallel}$-Fourier component of $\nabla_{\perp} U_p$ evaluated at the surface-mode wavevector as 
\begin{equation}
	S_{\bm{q}_{\parallel}} \propto \widetilde{\nabla_{\perp} U_p} (\bm{q}_{\parallel}=\bm{k}_{sp,\parallel}, \Omega=\omega_{sp}) \mathrm{,}
\end{equation}
where $\Omega$ is the temporal Fourier variable conjugate to the envelope of the laser pulse.
For a single monochromatic plane wave on a strictly infinite plane, $\widetilde{\nabla_{\perp} U_p} (\bm{q}_{\parallel}, \Omega=\omega_{sp})\propto \delta^{(2)}({\bm{q}_{\parallel}}) \delta(\Omega)$ and, thus, $S_{\bm{q}_{\parallel}}=0$ as the ponderomotive drive carries only zero in-plane momentum and therefore cannot excite a finite-$k_{sp,\parallel}$ surface mode on a perfectly translationally invariant plane. The translational constraint still holds for ponderomotive drive. 
On an infinite planar surface, the breaking of the translational symmetry relevant to the 1D envelope equation comes from the finite longitudinal envelope $L_0$ or any other modulation along the interface. A finite transverse spot $w_0$ modifies the normal overlap, but by itself does not supply a new tangential Fourier component along $z$.
In contrast, if the laser pulse is finite along the interface (for example, finite in $z$ as shown in Fig.~\ref{fig:geometry}), $U_p$ is no longer perfectly uniform tangentially, so $\widetilde{\nabla_{\perp} U_p}$ is not a delta function but has a finite envelope where the exact translational selection is relaxed.
While a finite transverse spot $w_0$ alone modifies the normal overlap, it does not by itself supply a new tangential Fourier component along $z$.

Let's consider a finite linearly polarised laser pulse of Gaussian profile, for example, which propagates parallel along the surface in $z$-direction at velocity $v_L$, 
\begin{equation}
	a(\xi_{\perp}, z, t) = a_0 e^{-(\xi_{\perp}-d)^2/w_0^2}  e^{-\zeta_L^2/L_0^2} \rm{,}
	\label{eq:Gaussian_envelope}
\end{equation}
where $\xi_{\perp}$ denotes the general transverse coordinate, e.g. $\xi_{\perp}=x$ in planar or $\xi_{\perp}=r$ in cylindrical geometry, respectively. $d$ denotes the distance from the laser centre to the axis. In the cylindrical case, $d=0$ where the laser propagates along the axis as shown in Fig.~\eqref{fig:geometry}. $\zeta_L=z-v_L t$. $w_0$ and $L_0$ are the laser transverse spot size and longitudinal length, respectively.
For LP, the cycle-averaged ponderomotive potential is $U_p = m_e c^2 (\gamma_L-1)$ with $\gamma_L(\xi_{\perp}, z, t)=\sqrt{1+s_p a^2(\xi_{\perp}, z, t)}$. Under the frozen-envelope approximation, $\gamma_L\simeq \gamma_q$, and 
\begin{equation}
\begin{split}
	\partial_{\perp} U_p \simeq 
	& - 2 s_pm_e c^2 \frac{a_0^2}{\gamma_q} \frac{\xi_{\perp}-d}{w_0^2}   \\
	& \cdot e^{-2(\xi_{\perp}-d)^2/w_0^2} e^{-2\zeta_L^2/L_0^2} \mathrm{.}	
\end{split}
\end{equation}
The overlap is
\begin{equation}
\begin{split}
	\mathcal{I}_{\perp} = \int d \mathbb{A} \, n_0 (\bm{r}_{\perp}) \bm{E}_{\rm{sp}, \perp}^{*}(\bm{r}_{\perp}) \cdot \hat{\bm{\xi}}_{\perp} \frac{d- \xi_{\perp}}{w_0^2} e^{-2(\xi_{\perp}-d)^2/w_0^2} \rm{,}
\end{split}
	\label{eq:overlap_general_1}
\end{equation}
where $\hat{\bm{\xi}}_{\perp}$ is the unit vector in the transverse direction along which the laser intensity varies away from the surface or axis.
The envelope equation is then given by
\begin{equation}
	(\partial_t + v_g \partial_z + \Gamma ) A = - i \frac{e c^2 s_p a_0^2}{2 \omega_{sp} \gamma_q^2\mathcal{N}_{sp}} \mathcal{I}_{\perp} e^{-2\zeta_L^2/L_0^2}  \mathrm{,}
	\label{eq:envelope_equ_full}
\end{equation}
which can be solved along characteristic $z-v_gt= \rm{const}$.  
It is seen from Eq.~\eqref{eq:overlap_general_1} that the surface geometry and extent modify the physical field in Eq.~\eqref{eq:physical_E_field} by changing the mode shape $\bm{E}_{\rm{sp}}$ and area element $d \mathbb{A}$, respectively.

\begin{figure}
	\centering
	\includegraphics[width=0.45\textwidth]{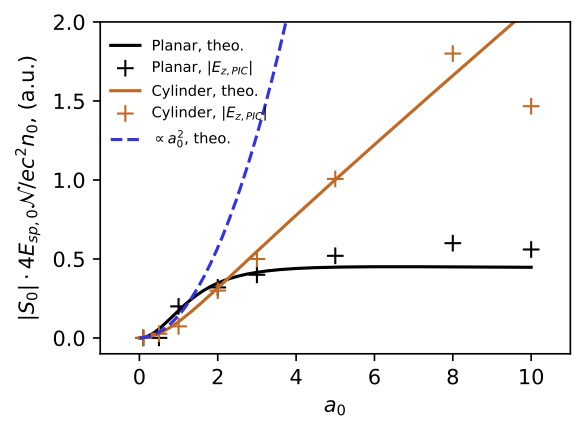}
	\caption{Overlap-normalised ponderomotive source coefficient $|S_0|$ calculated from Eq.~\eqref{eq:source_planar_general} for planar surface (black) and from Eq.~\eqref{eq:S0_cy_m} for cylindrical surface (orange). PIC markers show the corresponding on-surface electric-field amplitude $|E_{z, \rm PIC}|$. The theoretical source and PIC field are normalised to compare their $a_0$-scaling. 
	A quantitative field comparison requires solving Eq.~\eqref{eq:green_sol} and reconstructing $E_z=A E_{sp,z}$.
	The used parameters: $w_0=2d \approx 25 k_{\rm p}^{-1}$, $\zeta_{sp}=0$.}
	\label{fig:S0_normal}
\end{figure}

As seen from Eq.~\eqref{eq:envelope_equ_full}, for $a_0 \ll 1$, $S$ scales as $S \propto a_0^2$, as shown by the dashed blue line in Fig.~\ref{fig:S0_normal}, which is the familiar nonrelativistic ponderomotive scaling.
For the planar eigenmode evaluated here, the full overlap-normalised source saturates at large $a_0$, as shown by the black line. 
This is due to the relativistic modification of $\varepsilon(\omega)$, or, physically, the weaker confinement of the mode to the surface. 
A curved surface can effectively compensate for the saturation, as shown by the orange line, which clearly indicates both the geometric and relativistic effects.
This feature will enable the nonlinear plasma wakefield generation~\cite{Lei:2025aa, Lei:2025ab}. 
Fully 3D PIC  simulations are carried out with the code WarpX~\cite{vay_2025_17261711}, which agrees well with the theoretical predictions before the surface becomes soft, as shown in Fig.~\ref{fig:S0_normal}. With a high $a_0$, such as $a_0>5$, the strong pulse front can result in the surface electrons expanding into the vacuum with a density gradient and weakening the field confinement, which is referred to as a softened surface here. This effect should be further studied for practical applications with high-power lasers.
It also shows in PICs that the field amplitude on a cylindrical surface is one order of magnitude higher than that on a planar surface. 
The PIC configuration is as follows: dimensions of the moving window are $8~\si{\mu m} \times 8~\si{\mu m} \times 30~\si{\mu m}$, comprising $256\times 256 \times 512$ cells in the $x$, $y$, and $z$ directions, respectively. Each cell contains 8 macro particles.
The planar and cylindrical surface geometries used in PICs are shown in Fig.~\ref{fig:geometry}. 
The initial surface electron density is $n_0=1\times 10^{22}~\si{cm^{-3}}$. The laser spot size is $2~\si{\mu m}$ and duration $27~\si{fs}$. In the planar surface, $d=1~\si{\mu m}$. In the cylindrical surface, the radius $R=1~\si{\mu m}$. This indicates the on-surface laser strength is $0.78 a_0$.


The general solution Eq.~\eqref{eq:imhomo_wave_equ_slowenv_full} can be easily found by using Green's method. 
Let $\zeta=z-v_g t$, Eq.~\eqref{eq:imhomo_wave_equ_slowenv_full} becomes a 1D inhomogeneous ordinary differential equation (ODE)
\begin{equation}
	\frac{d}{dt} A(\zeta+v_g t, t) + \Gamma A(\zeta + v_g t, t) =  S(\zeta+v_g t, t) \mathrm{,}
	\label{eq:green_sol}
\end{equation}
with the solution 
\begin{equation}
	A(z,t) = \int_{-\infty}^t dt' \, e^{-\Gamma(t-t')} S(z-v_g(t-t'), t') \rm{.}
	\label{eq:solv_1D_ODE}
\end{equation}
The solution in Eq.~\eqref{eq:solv_1D_ODE} is exact for Eq.~\eqref{eq:imhomo_wave_equ_slowenv_full}. It needs no boundary assumptions, as no SP is before the driver arrives. 

Considering a Gaussian envelope drive as in Eq.~\eqref{eq:Gaussian_envelope}, the source term is 
\begin{equation}
	S(z,t) = S_0 e^{-2\zeta_L^2/L_0^2} \mathrm{,}
    \label{eq:S0_planar_gaussian}
\end{equation}
where $S_0$ is a geometry-dependent constant 
\begin{equation}
\begin{split}
	S_0 = - i \frac{e c^2 s_p a_0^2}{2 \omega_{sp} \gamma_q^2 \mathcal{N}_{sp}} \int d \mathbb{A} \, & n_0 (\bm{r}_{\perp}) \bm{E}_{\rm{sp}, \perp}^{*} (\bm{r}_{\perp}) \cdot \hat{\bm{\xi}}_{\perp} \\
	& \frac{d - \xi_{\perp}}{w_0^2} e^{-2(\xi_{\perp}-d)^2/w_0^2} \rm{.} \nonumber
\end{split}	
\end{equation} 
Eq.~\eqref{eq:S0_planar_gaussian} denotes the resonant source envelope after projection onto the RSP phase. If the full non-resonant moving Gaussian source is retained instead, an additional detuning factor must be included, as in the direct electric-field drive solution.

\begin{figure}
	\centering
	\includegraphics[width=0.45\textwidth]{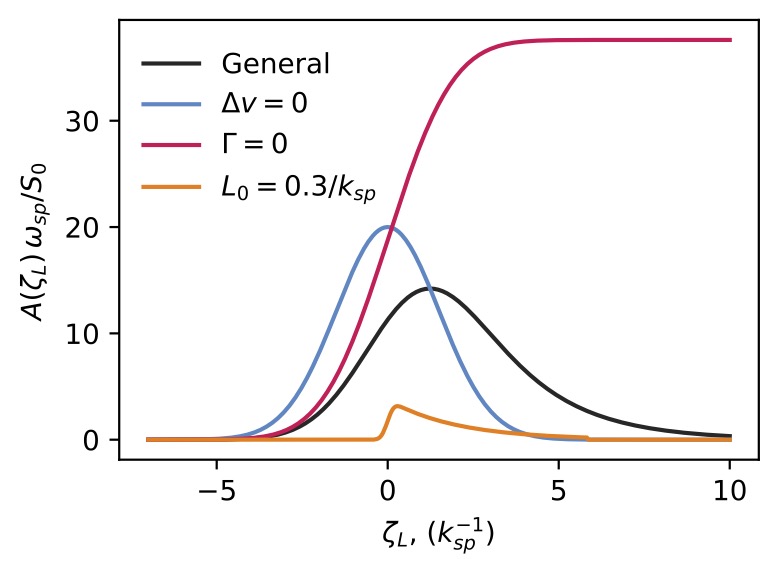}
	\caption{Envelope $A$ as a function of $\zeta_L$ for general case in Eq.~\eqref{eq:A_solve_erfc} (red), $\Delta v=0$ (blue), $\Gamma=0$ (black) and $L_0=0.3/k_{sp}$. In general case, the used parameters: $\Delta v = 0.1c$, $\Gamma=0.05 \omega_{sp}$, $L_0=3/k_{sp}$, which are changed in each specific case. The laser propagates to the left.}
	\label{fig:envelope_general}
\end{figure}

The solution in Eq.~\eqref{eq:solv_1D_ODE} is written as
\begin{equation}
	A(z,t) = S_0 \int_{-\infty}^{t} dt'\, e^{-\Gamma (t-t')} \exp\left \{- \frac{2 \left [\zeta_L - \Delta v (t-t') \right ]^2}{L_0^2} \right \} \rm{,} \nonumber
	\label{eq:A_gaussian_general}
\end{equation}
where we denote $\Delta v = v_g - v_L$. 
Under the frozen-envelope approximation used in Eq.~\eqref{eq:envelope_electric_general}, $S_0$ is constant and the Gaussian source admits a closed-form solution in terms of the complementary error function $\rm{erfc}$ as
\begin{equation}
\begin{split}
	A(\zeta_L) = & \frac{S_0}{2} \sqrt{\frac{\pi}{2}} \frac{L_0}{|\Delta v|} \exp\left[\frac{\Gamma}{\Delta v} \left( -\zeta_L + \frac{\Gamma L_0^2}{8 \Delta v} \right) \right] \\
	& \cdot \mathrm{erfc}\left(\frac{\Gamma L_0}{2 \sqrt{2} |\Delta v|} - \mathrm{sgn} (\Delta v) \frac{\sqrt{2} \zeta_L }{L_0} \right) \mathrm{,}
	\label{eq:A_solve_erfc}
\end{split}
\end{equation}
which is shown in Fig.~\ref{fig:envelope_general} for different cases.
If the source is not already projected onto the resonant RSP phase, the source term contains an additional phase factor and the solution acquires a detuning parameter analogous to Eq.~\eqref{eq:electric_full_Am}.

With the velocity-matched drive, $\Delta v \to 0$ and $\Gamma > 0$, Eq.~\eqref{eq:A_solve_erfc} presents a steady-source solution as 
\begin{equation}
	A(z, t) = \frac{S_0}{\Gamma} \exp \left[- \frac{2\zeta_L^2}{L_0^2} \right ] \mathrm{,}
\end{equation}
which is the steady matched-source result implied by Eq.~\eqref{eq:solv_1D_ODE} with the lower limit $t' =- \infty$.
This means that the local source seen by the mode is constant in time. Therefore, the envelope builds a steady value set by a balance between the drive and loss  $S_0/\Gamma$. If $\Gamma=0$, the integral diverges, which presents an unbounded resonant pumping.

Without loss, e.g. $\Gamma = 0$ and $\Delta v \neq 0$, Eq.~\eqref{eq:A_solve_erfc} can be simplified as
\begin{equation}
	A(z,t) = \frac{S_0}{2} \sqrt{\frac{\pi}{2}} \frac{L_0}{|\Delta v|} \left[1 + \mathrm{erf} \left( \mathrm{sgn}(\Delta v) \frac{\sqrt{2} \zeta_L}{L_0} \right) \right] \rm{,}
\end{equation}
which diverges if $\Delta v =0$.
Far ahead of the driver, $\mathrm{sgn}(\Delta v) \zeta_L \ll - L_0$, and then $A \to 0$. For behind $\mathrm{sgn}(\Delta v) \zeta_L \gg L_0$, 
\begin{equation}
	A \to S_0  \sqrt{\frac{\pi}{2}} \frac{L_0}{|\Delta v|}
\end{equation}
which gives the amplitude of the wakefield.

In the short-drive limit, $L_0 \to 0$, the RSP envelope jumps when the laser centre passes the point. Eq.~\eqref{eq:A_solve_erfc} becomes
\begin{equation}
	A(z,t) \to  S_0 \sqrt{\frac{\pi}{2}} \frac{L_0}{|\Delta v|} e^{-\Gamma \zeta_L / \Delta v} H \bigg (\frac{\zeta_L}{\Delta v} \bigg) \rm{,}
\end{equation}
where $H$ is the Heaviside step.
With no loss, $\Gamma=0$, and then $e^{-2\zeta_L^2/L_0^2} \to (\sqrt{\pi/2}) L_0 \delta(\zeta_L)$, Eq.~\eqref{eq:A_solve_erfc} becomes
\begin{equation}
		A(z,t) \to  S_0 \sqrt{\frac{\pi}{2}} \frac{L_0}{|\Delta v|} H \bigg (\frac{\zeta_L}{\Delta v} \bigg) \rm{,}
\end{equation}
which is a pure step set by the impulse area. Noted that exactly $L_0=0$ gives zero area.

In the exactly matched case $v_g=v_L$ or $\Delta v=0$, the laser coordinate $\zeta_L$ is constant along the characteristic. The source then becomes time-independent along that characteristic, so the envelope grows secularly in the absence of loss as 
\begin{equation}
	A(z,t) = S_0 \int_{t_i}^{t} dt' \, e^{-2\zeta_L^2/L_0^2} = S_0 (t-t_i) e^{-2\zeta_L^2/L_0^2}
\end{equation}
where $t_i$ denotes the initial time. 
In practice, we need loss ($\Gamma > 0$), group velocity mismatch ($\Delta v = v_g - v_L \neq 0$), ponderomotive detuning $\Delta_p \neq 0$, or a finite interaction length (e.g. laser-SP overlap exists only over a finite time) to terminate the resonant pumping. 

In the following section, we provide the explicit solutions to the RSP excitation on planar and cylindrical surfaces driven by a Gaussian pulse, based on the theory developed above.

\subsection{On planar surface}

The planar geometry is shown in Fig.~\ref{fig:geometry}(a). The vacuum region is at $x>0$ while the plasma is at $x<0$.
For an isotropic, nonmagnetic plasma-vacuum interface, the RSP is TM-polarised where $\bm{E}_{\mathrm{sp}}=(E_{\mathrm{sp},x}, 0, E_{\mathrm{sp},z})$ and $\bm{B}_{\rm{sp}}=(0, B_{\mathrm{sp},y},0)$. 
The eigen equation becomes 
\begin{equation}
	(\frac{d^2}{dx^2} - \kappa_j^2) E^{(j)}_z(x)=0 \mathrm{,}
\end{equation}
where the decay parameters are defined as
\begin{equation}
	\kappa_j(\omega, k_z) = \sqrt{k_z^2 - \varepsilon_j(\omega) k_0^2} \mathrm{,}
	\label{eq:kappa_j}
\end{equation}
and $j$ is either 1 or 2, denoting the solutions in vacuum or plasma, respectively. $k_0$ denotes the vacuum wavenumber. For the bound surface mode, we need $\Re\kappa_j >0$.
The eigen electric fields can be obtained by solving the homogeneous wave equation Eq.~\eqref{eq:driven_wave_equ} in Cartesian coordinates $(x,y,z)$ in vacuum $x>0$ 
\begin{equation}
		E_{z}^{(1)} (x) = E_{0} e^{-\kappa_{1} x}
		\label{eq:eigen_ez_pl_1}
\end{equation}
and in plasma  $x<0$
\begin{equation}
	E_{z}^{(2)} (x) = E_{0} e^{\kappa_{2} x} \mathrm{.} 
	\label{eq:eigen_ez_pl_2}
\end{equation}
The dielectric function is $\varepsilon_1=1$ in vacuum and $\varepsilon_2(\omega)=1-\omega_p^2/\gamma_q \omega^2$.
Other components are given from Maxwell's equations as 
\begin{equation}
	E_{x}^{(1)} = \frac{ik_z}{\kappa_{1}} E^{(1)}_{z} \quad \text{and} \quad E_{x}^{(2)} = -\frac{ik_z}{\kappa_{2}} E^{(2)}_{z} \mathrm{,}
\end{equation}
and
\begin{equation}
	B_{y}^{(1)} = \frac{i\omega \varepsilon_1}{c \kappa_{1}} E^{(1)}_{z}  \quad \text{and} \quad 	B_{y}^{(2)} = -\frac{i\omega \varepsilon_2}{c \kappa_{2}} E^{(2)}_{z} 
\end{equation}

For the TM surface wave ($p$-polarisation), the dispersion function can be obtained from the boundary condition in Eq.~\eqref{eq:boundary_cross} and ~\eqref{eq:boundary_normal}, which give
\begin{equation}
	\mathcal{D}_{\rm pl} (\omega, k_z) = \frac{\varepsilon_1}{\kappa_1 } + \frac{\varepsilon_2}{\kappa_2 }  \mathrm{,}
	\label{eq:disp_pl_1}
\end{equation}
where the eigenmode satisfies $\mathcal{D}_{\rm pl} (\omega_{sp},k_{sp})=0$ or in the well-known form 
\begin{equation}
	k_z = k_0 \sqrt{ \frac{\varepsilon_1\varepsilon_2}{\varepsilon_1+ \varepsilon_2}} \mathrm{.}
	\label{eq:disp_pl}
\end{equation}
A TM surface-mode solution requires opposite-sign permittivities, but a bound surface plasmon additionally requires  $\Re{\kappa_1}>0$ and $\Re{\kappa_2}>0$. For the usual lossless vacuum-plasma branch with $\varepsilon_1=1$, this implies $\varepsilon_2<-1$. Therefore, the bound RSP exists only when the plasma permittivity is sufficiently negative. This is exactly the case for metal-dielectric or plasma vacuum interface because $\varepsilon_{p}=1-\omega_p^2/\omega^2 <-1$ when $\omega<\omega_p/\sqrt{2\gamma_q}$. 
Therefore, RSP is a legitimate Maxwell eigenmode.


The phase velocity is $v_{\rm ph}=\omega/k_z$ and group velocity $v_g=-(\partial \mathcal{D}_{\rm pl}/\partial k_z) / (\partial \mathcal{D}_{\rm pl}/ \partial \omega)$ which indicates that the relativistic effect lowers the phase and group velocities.
As seen from Eq.~\eqref{eq:disp_pl}, the bound RSP requires $\varepsilon_2<-1$, which indicates the lower limit of plasma density for the RSP solution to exist 
\begin{equation}
	n_0 > 2 \gamma_q n_c(\omega_{sp}) \mathrm{,}
	\label{eq:lowdensity_limit}
\end{equation}
where $n_c(\omega_{sp})=m_e \omega_{sp}^2/4 \pi e^2$ is the critical density evaluated at the RSP frequency.
In the large-$k$ electrostatic limit $k \to \infty$, $\varepsilon_2 \to -1$. As a result, we can get the resonant frequency
\begin{equation}
	\omega_{sp,\max} = \frac{\omega_p}{\sqrt{2 \gamma_q}} \mathrm{,}
	\label{eq:asy_freq}
\end{equation}
which is the maximum frequency that an RSP can obtain.
In this resonant case, wavelength $\lambda=2\pi/k\to 0$ and the group velocity $v_g = \partial \omega /\partial k \to 0$. So the wave becomes extremely compressed along the surface. This corresponds to very strong field localisation. The wave becomes almost electrostatic physically. SPP becomes essentially a pure charge oscillation where retardation effects disappear. 
This is important for strong-field surface plasmon physics, which enables relativistic electron motion and strong surface currents.
The divergence of the planar surface-mode wavevector in the electrostatic limit is a feature of the ideal local, lossless, infinitely sharp planar model with exact translational symmetry. In practice, nonlocal response, finite density gradients, dissipation, finite size, and curvature all regularise this behaviour.

For TE surface wave ($s$-polarisation), we can similarly obtain the dispersion relation as
\begin{equation}
	\frac{\kappa_1}{\mu_1} + \frac{\kappa_2}{\mu_2 } = 0 \mathrm{.}
\end{equation}
Now, the existence condition is $\mu_1 \mu_2 <0$, meaning the magnetic permeabilities must have opposite signs.  However, for almost all natural materials, $\mu \approx 1>0$. Thus, the TE surface wave condition cannot be satisfied.  An unmagnetised plasma cannot support s-polarised surface waves. 
The TE surface-wave condition can be satisfied in systems with an effective negative magnetic response, including single-negative magnetic metamaterials~\cite{Smith:2004aa, Hotta:2004aa}, magnetised ferrite~\cite{Ali:2018aa} or gyromagnetic media~\cite{Macedo:2019aa}, and photonic-crystal structures~\cite{Vinogradov:2006aa,Huang:2004aa}.


%

From Eq.~\eqref{eq:S0_planar_gaussian}, the source term is calculated on planar surface, e.g. $d \mathbb{A}=dx$ and $\xi_{\perp} = x$, as
\begin{equation}
\begin{split}
	S_{0,\mathrm{pl}} = -i \frac{e c^2 s_p a_0^2}{2 \omega_{sp} \gamma_q^2 \mathcal{N}_{sp}} &
	 \int_{-\infty}^{0} dx \, n_0(x) E^{(2) *}_{sp,x}(x) \\
	& \cdot \frac{d-x}{w_0^2} e^{-2(x-d)^2/w_0^2} \mathrm{.}
		\label{eq:source_planar_general}
\end{split}
\end{equation}
Eq.~\eqref{eq:source_planar_general} can be analytically integrated for a small spot size $\kappa_2 w_0 \ll 1$ as
\begin{equation}
	S_{0,\mathrm{pl}} \approx -i \frac{e c^2 s_p a_0^2 n_0}{8 \omega_{sp} \gamma_q^2 \mathcal{N}_{sp}}  E^{(2)*}_{x}(0^{-}) e^{-2d^2/w_0^2} 
\end{equation}
and for a wide spot size $\kappa_2 w_0 \gg 1$,
\begin{equation}
	S_{0,\mathrm{pl}} \approx -i \frac{e c^2 s_p a_0^2 n_0}{2 \omega_{sp} \gamma_q^2 \mathcal{N}_{sp} }  E^{(2)*}_{x}(0^{-}) \frac{d \kappa_2 + 1}{\kappa_2^2 w_0^2} e^{-2d^2/w_0^2} \mathrm{,}
\end{equation}
where $E^{(2)*}_{x}(0^{-})$ represents the plasma-side normal electric field of the SP mode evaluated right at the interface. 

\subsection{On cylindrical surface}
The cylindrical geometry is shown in Fig.~\ref{fig:geometry}(b). The radius of the solid tube is $R$. The vacuum region is inside the cylinder $r<R$ and plasma $r>R$. The laser propagates along the axis $r=0$.
The eigen equation becomes 
\begin{equation}
	\left [ \frac{1}{r}\frac{d}{dr}\left( r \frac{d}{dr}\right ) - \frac{m^2}{r^2}- \kappa_j^2 \right ] \tilde{E}^{(j)}_{z,m} (r) = 0 \mathrm{,}
	\label{eq:vHelmholtz_ope}
\end{equation}
where $m$ represents the azimuthal index. $\kappa_1$ and $\kappa_2$ are defined in Eq.~\eqref{eq:kappa_j}. 
$\tilde{E}^{(j)}_{z,m} (r)$ represents the radial profile.
Eq.~\eqref{eq:eigenmodes_equs} becomes the modified Bessel equation. 
Eq.~\eqref{eq:vHelmholtz_ope} shows that the effective tangential wavenumber becomes $k^2_{\parallel} = k_z^2 + m^2/R^2$ near the cylindrical surface $r\simeq R$. So the cylindrical curvature removes the continuous 2D translational symmetry by replacing one continuous tangential momentum component with a discrete azimuthal index $k_{\phi} \to m/R$. This discretisation implies that an axisymmetric source, such as the ponderomotive drive of the on-axis pulse in Fig.~\ref{fig:geometry}(b), couples only to the $m=0$ cylindrical mode.

Similar to the planar case, for the axisymmetric cylindrical mode $m=0$, only the TM modes exist.
For $m\neq 0$, the exact electromagnetic cylindrical modes are generally hybrid. Nevertheless, in the strongly surface-confined or electrostatic limit, the TM-dominant part can be described by the longitudinal electric field as 
\begin{equation}
	\begin{split}
		& \tilde{E}_{z,m}^{(1)}(r) = E_{ 0} I_m(\kappa_{1} r)  \\
		& \tilde{E}_{z,m}^{(2)}(r) = E_{ 0} \frac{I_m(\kappa_{1} R)}{K_m(\kappa_{2} R)} K_m(\kappa_{2} r) \mathrm{,}\\
	\end{split}
	\label{eq:eigenmodes_cyl}
\end{equation} 
with the full field written as
\begin{equation}
	E_{z,m}^{(j)} = \tilde{E}_{z,m}^{(j)}(r)e^{im\phi + i k_z z - i \omega t}
\end{equation}
The evanescent nature of the SP field is presented by the modified Bessel functions $I_m$ and $K_m$ in vacuum and plasma, respectively. 
Notably, the $m=0$ mode does not decay to zero in the vacuum channel region, but rather provides an on-axis acceleration field. This feature is significant for RSP-based wakefield acceleration.

Other components are given by
\begin{equation}
	\begin{split}
		& \tilde{E}_{r,m}^{(j)} = -\frac{i k_z}{\kappa^2_{j}} \partial_r \tilde{E}_{z,m}^{(j)} \mathrm{,} \quad  \tilde{E}_{\phi,m}^{(j)} = \frac{mk_z}{\kappa_j^2 r} \tilde{E}^{(j)}_{z,m} \\
		& \tilde{B}_{r,m}^{(j)} = - \frac{\omega \varepsilon_j m}{c\kappa_j^2 r} \tilde{E}^{(j)}_{z,m} \mathrm{,} \quad \tilde{B}_{\phi,m}^{(j)} = - \frac{i \omega \varepsilon_j }{c \kappa^2_{j}} \partial_r \tilde{E}_{z,m}^{(j)} \mathrm{.}
	\end{split}
\end{equation}

Similar to the planar case, the source can be calculated on a cylindrical surface, $d \mathbb{A}=rdr d \phi$ and $\xi_{\perp} = r$, as
\begin{equation}
	\begin{split}
		S_{0,cy,m} = & i \frac{e c^2 s_p a_0^2}{2\omega_{sp} \gamma_q^2 \mathcal{N}_{sp,m}} \int_{0}^{2\pi} d \phi  \, e^{-im\phi}\int_{R}^{\infty} r dr \, \\
		& \cdot n_0(r) \tilde{E}^{*}_{sp, r, m} (r)  \frac{r}{w_0^2} e^{-2r^2/w_0^2} \mathrm{,}
		\label{eq:S0_cy_m}
\end{split}
\end{equation}
For $m=0$ mode,
\begin{equation}
\begin{split}
		S_{0,cy,0} = & i \frac{\pi e c^2 s_p a_0^2}{\omega_{sp}\gamma_q^2 \mathcal{N}_{sp,0}} \, \int_{R}^{\infty} r dr \, n_0(r)  \\
		& \cdot  \tilde{E}^*_{sp,r,0}(r) \frac{r}{w_0^2} e^{-2r^2/w_0^2} \mathrm{,}
\end{split}
\end{equation}
and the envelope is then calculated by Eq.~\eqref{eq:solv_1D_ODE}.
Therefore, an axisymmetric ponderomotive drive excites only the $m=0$ cylindrical mode, while all $m\neq 0$ components vanish by azimuthal orthogonality.

\begin{figure}
	\centering
	\includegraphics[width=0.45\textwidth]{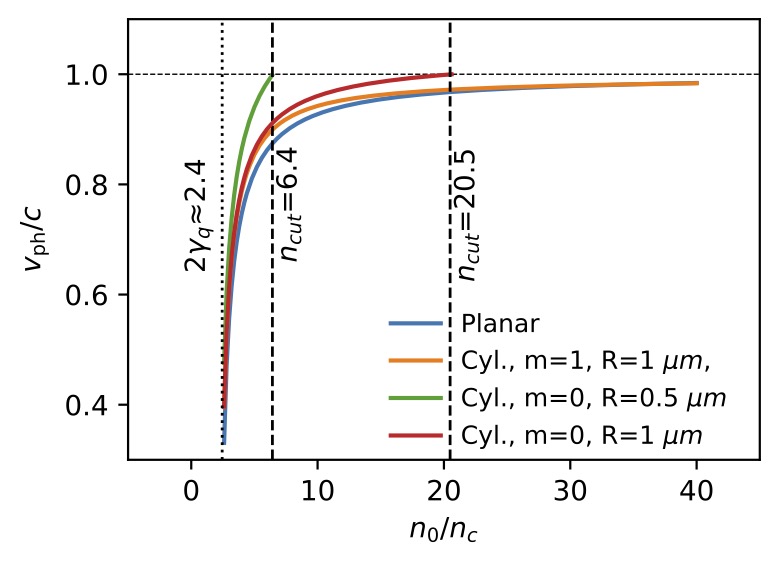}
	\caption{(a) Phase velocity $v_{ph}/c$ as a function of surface electron density $n_0/n_c$ on a planar interface (blue) calculated from Eq.~\eqref{eq:disp_pl} and on a cylindrical interface for $m=1$ (orange) and $m=0$ (green and red) modes calculated from Eq.~\eqref{eq:disp_cyl}, respectively. The vertical dotted lines show the density cutoff from Eq.~\eqref{eq:lowdensity_limit} and the dashed lines from \eqref{eq:n_cutoff} with $R=0.5~\si{\mu m}$ or $R=1~\si{\mu m}$. The horizontal dashed line shows the light line. The laser pulse is linearly polarised with normalised strength $a_0=1.0$ and wavelength $\lambda_L=0.8~\si{\mu m}$.}
	\label{fig:disp_cyl}
\end{figure}

According to the continuity of $E_{\mathrm{sp},z}$ and $B_{\mathrm{sp},\phi}$, the dispersion function is given as
\begin{equation}
	\mathcal{D}_{\mathrm{cy},m}(\omega, k_z) = \frac{\varepsilon_1}{\kappa_1} \frac{I_m^{\prime}(\kappa_1 R)}{I_m(\kappa_1 R)} - \frac{\varepsilon_2}{\kappa_2} \frac{K_m^{\prime}(\kappa_2 R)}{K_m(\kappa_2 R)} \mathrm{,}
	\label{eq:disp_cyl}
\end{equation}
where the eigenmodes satisfy $\mathcal{D}_{\mathrm{cy},m}(\omega_{sp},k_{sp})=0$.
For $m \neq 0$, the exact electromagnetic cylindrical surface modes are generally hybrid. In that case, Eq.~\eqref{eq:disp_cyl} should be regarded as the TM-dominant or electrostatic approximation. The exact dispersion follows from the full hybrid boundary-condition determinant~\cite{Little:2012aa, Lei:2025oty}.

Here, the curvature enters directly through the Bessel structure and through the discrete label $m$.
For $\kappa_j R\gg 1$ and modest $m$, Eq.~\eqref{eq:disp_cyl} reduces to the planar condition in Eq.~\eqref{eq:disp_pl_1}. 
The phase velocity $v_{\rm ph}$ is calculated from Eq.~\eqref{eq:disp_cyl} and compared with that on the planar interface as shown in Fig.~\ref{fig:disp_cyl}. It is seen that the curved surface introduces two principal effects by modifying the wavenumber $k_{sp}$, including  1) shifts in $v_{\rm ph}$, and 2) $m=0$ mode high-density cutoff.
The curvature-induced reduction of $k_{sp}$ increases $v_{\rm ph}$ and enlarges the accessible low-density range, which is beneficial for relativistic particle acceleration.
\textit{a. Dispersion shift:} When the cylinder is gently curved, i.e. $\kappa_{j} R \gg 1$, using the expansions
\begin{equation}
	\frac{I^{'}_m}{I_m} = 1 - \frac{1 }{2 \kappa_1 R} + \mathcal{O}\left( (\kappa_1 R)^{-2}\right ) \rm ,
	\label{eq:ImImp}
\end{equation}
and 
\begin{equation}
	\frac{K^{'}_m}{K_m} = -1 - \frac{1 }{2 \kappa_2 R} + \mathcal{O}\left( (\kappa_2 R)^{-2}\right ) \rm ,
\end{equation}
the dispersion in Eq.~\eqref{eq:disp_cyl} can be written by
\begin{equation}
	\mathcal{D}_{\mathrm{cy},m} = \mathcal{D}_{\rm pl}- \frac{1}{2 R} \left( \frac{\varepsilon_1}{\kappa_1^2} - \frac{\varepsilon_2}{\kappa_2^2}\right) + \mathcal{O}\left( (\kappa_{1,2} R)^{-2} \right) \mathcal{,}
	\label{eq:disp_cy_expand} 
\end{equation}
where the second term presents the curvature effect and enters at order $1/R$.
As the first term $\mathcal{D}_{\rm pl}$ is exactly the planar dispersion function in Eq.~\eqref{eq:disp_pl_1}, the cylindrical dispersion can be written in the form of $\mathcal{D}_{\rm pl} + \delta \mathcal{D}_{\rm cy} =0$ where
\begin{equation}
	\delta \mathcal{D}_{\rm cy} = - \frac{1}{2 R} \left( \frac{\varepsilon_1}{\kappa_1^2} - \frac{\varepsilon_2}{\kappa_2^2}\right) \mathrm{,}
	\label{eq:delta_Dcy}
\end{equation}
introducing the shifts in phase and group velocity.
At the order retained in Eqs.~\eqref{eq:ImImp}-\eqref{eq:delta_Dcy}, the curvature correction is independent of $m$. The explicit m-dependence enters at the next order in $(\kappa_j R)^{-1}$, or through the full Bessel-function dispersion relation in Eq.~\eqref{eq:disp_cyl}.
The shift of phase velocity $v_{\rm ph}$ can be estimated by the shift of wavenumber $\delta k_z$ and frequency $\delta \omega$ as
\begin{equation}
	\frac{\delta v_{\rm ph}}{v_{\rm ph}} \approx \frac{\delta \omega}{\omega} - \frac{\delta k_z }{k_z} \mathrm{.}
	\label{eq:shift_v_ph}
\end{equation}
Shift of $k_z$ can be calculated at fixed $\omega$, 
\begin{equation}
	\delta k_z \bigg |_{\omega} = - \frac{\delta \mathcal{D}_{\rm cy}}{\partial_{k_z} \mathcal{D}_{\rm pl} } \mathrm{,}
	\label{eq:shift_k}
\end{equation}
where 
\begin{equation}
	\frac{\partial \mathcal{D}_{\rm pl}}{\partial k_z} = - k_z \left(\frac{1}{\kappa_1^3} + \frac{\varepsilon_2}{\kappa_2^3} \right ) \mathrm{.} \nonumber
\end{equation}
Shift of $\omega$ at fixed $k_z$ is
\begin{equation}
	\delta \omega \bigg |_{k_z} = - \frac{\delta \mathcal{D}_{\rm cy}}{\partial_{\omega} \mathcal{D}_{\rm pl}} \mathrm{,}
	\label{eq:shit_omega}
\end{equation}
where 
\begin{equation}
	\frac{\partial \mathcal{D}_{\rm pl}}{\partial \omega} = \frac{\partial \varepsilon_2 / \partial \omega}{\kappa_2} + \frac{\varepsilon_2}{2 c^2 \kappa_2^3} \left( \omega^2 \frac{\partial \varepsilon_2}{\partial\omega} +2 \omega \varepsilon_2  \right ) + \frac{\omega}{c^2 \kappa_1^3} \mathrm{.} \nonumber
\end{equation}

From Eq.~\eqref{eq:shift_v_ph}, \eqref{eq:shift_k} and \eqref{eq:shit_omega}, it is easy to see that for $m=0$ mode at fixed $\omega$, the correction only comes from $\delta k$ in Eq.~\eqref{eq:shift_k} where $\delta k <0$. As a result, the phase velocity is faster than that in the planar case, which scales as $\mathcal{O}(1/R)$. 
This feature is significant for the phase matching and wakefield acceleration.
For $m\neq 0$ mode, the azimuthal variation introduces an azimuthal surface wavenumber $k_{\phi}=m/R$ and the tangential wavenumber is replaced by $k_{\parallel}$. At a fixed $\omega$, $v_{\rm ph}$ decreases with $m$.
Similarly, Eq.~\eqref{eq:shift_k} and \eqref{eq:shit_omega} also show that curvature perturbs the group velocity $v_g=\partial \omega / \partial k_z$ via the same scale $\mathcal{O}(1/R)$.

\textit{b. High-density cutoff in $m=0$ mode:} 
From Eq.~\eqref{eq:disp_cyl}, it is easy to see that for $m=0$, at the limit $k_z \to k_0^{+} = \omega /c^{+}$ (just above the light line), $\kappa_1 \to 0$, 
\begin{equation}
	\frac{1}{\kappa_1} \frac{I_0^{\prime}(\kappa_1 R)}{I_0(\kappa_1 R)} \to \frac{R}{2}
\end{equation}
and then 
\begin{equation}
\begin{split}
		\mathcal{D}_{\mathrm{cy},0}(k_0^{+}) & = \frac{\varepsilon_1}{\kappa_1} \frac{I_0^{\prime}(\kappa_1 R)}{I_0(\kappa_1 R)} - \frac{\varepsilon_2}{\kappa_2} \frac{K_0^{\prime}(\kappa_2 R)}{K_0(\kappa_2 R)} \\
		& \approx \frac{ R}{2} + \frac{\varepsilon_2}{\kappa_2}\frac{K_1(\kappa_2 R)}{K_0(\kappa_2 R)}  \mathrm{,}
\end{split}
\end{equation}
where the first term is finite and the second term is negative as $\varepsilon_2 < 0$.
Therefore, the sign of $D(k_0^{+})$ depends on $\varepsilon_2$ and $R$. Only if $D(k_0^{+})$ is positive, there is an RSP solution. This creates a geometric existence condition for the $m=0$ cylinder mode that does not exist in other $m\geq 1$ modes and the planar case.
The curvature removes the guaranteed divergence that the planar dispersion has, and $m=0$ is the only azimuthal index where the vacuum-core term stays finite at the light line. Therefore, a solution may or may not exist depending on how negative $\varepsilon_2$ is, which depends on $n_0$, and the radius $R$.

\begin{figure}
	\centering
	\includegraphics[width=0.45\textwidth]{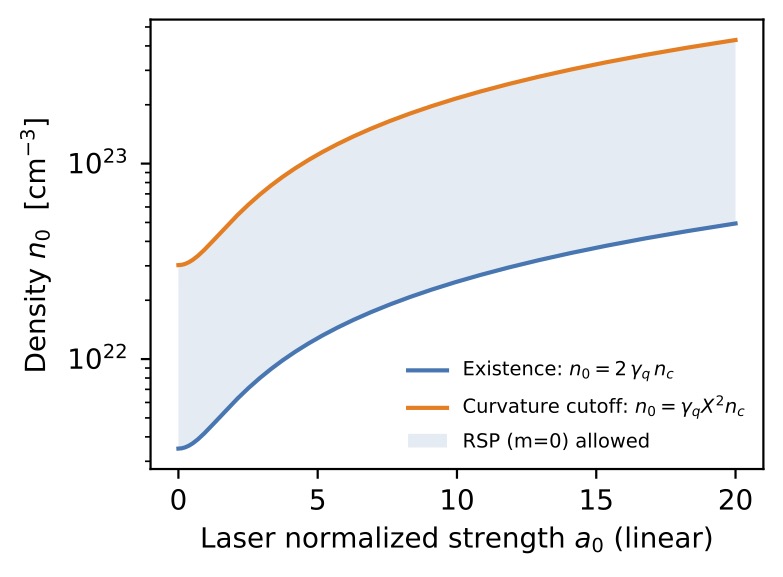}
	\caption{Density range for $m=0$ RSP mode to exist as a function of laser strength $a_0$ driven by an LP laser pulse.}
	\label{fig:highdensity_cutoff}
\end{figure}

For the $m=0$ cylindrical mode, the light-line limit yields a finite vacuum-core contribution, unlike the planar case and the $m \geq 1$ cylindrical modes.  Requiring \begin{equation}
	\mathcal{D}_{\rm cy,0}(k_0^{+}) > 0
\end{equation} 
or 
\begin{equation}
\frac{R}{2} > \frac{|\varepsilon_2|}{\kappa_2} \frac{K_1(\kappa_2 R)}{K_0(\kappa_2 R)} \mathrm{.}
\label{eq:cutoff_condition}
\end{equation}
provides a necessary light-line condition for the existence of a bound $m=0$ mode. 
In the large $\kappa_2 R$ limit $\kappa_2 R \gg 1$, this yields the approximate cutoff estimate 
\begin{equation}
	n_{\rm cut} \approx \gamma_q \chi^2 n_c \mathrm{,}
	\label{eq:n_cutoff}
\end{equation}
where 
\begin{equation}
	\chi = \frac{\pi R}{2 \lambda_{sp}} + \frac{1}{2} \sqrt{\frac{\pi^2 R^2}{\lambda_{sp}^2} + 4} \rm ,
\end{equation}
where $\lambda_{sp}=2\pi c/\omega_{sp}$.
Eq.~\eqref{eq:n_cutoff} is a useful asymptotic estimate of the upper density boundary. The actual cutoff should be obtained by solving the full root condition $\mathcal{D}_{\rm cy, 0}(\omega, k_z)=0$.
Combined with the bound-mode threshold in Eq.~\eqref{eq:lowdensity_limit}, this gives the approximate density window
\begin{equation}
	2 \gamma_q < n_0/n_c < \gamma_q \chi^2 \mathcal{,}
	\label{eq:density_cut_m0}
\end{equation}
which is shown in Fig.~\ref{fig:highdensity_cutoff}. This condition is important for high-power laser-driven RSP excitation. 
As Fig.~\ref{fig:disp_cyl} shows, the available density band is narrower for the $m=0$ mode on a more curved interface in the relativistic regime. This requires closer attention in experiments. 

\section{With electric field drive}
RSP can be directly driven by an external electric field, such as a laser field, $\bm{E}_L$. 
The external force acting on electrons is $\bm{f}_{\rm ext}=-e \bm{E}_L$.
The corresponding source current in Eq.~\eqref{eq:ext_current} becomes
\begin{equation}
	\bm{J}_{\rm ext}^{(L)} = i \frac{e^2n_0}{\omega_L m_e \gamma_q} \bm{E}_L \mathrm{,}
\end{equation}
where $\omega_d=\omega_L$.
The envelope equation in Eq.~\eqref{eq:imhomo_wave_equ_slowenv_full_2} then becomes 
\begin{equation}
\begin{split}
	(\partial_t + v_{g,m} \partial_z + \Gamma_m ) A_m = 
	& -\frac{i e^2}{4 \omega_L m_e \gamma_q \mathcal{N}_{sp,m}} \int d \mathbb{A} \, n_0(\bm{r}_{\perp})  \\
	& \bm{E}^{*}_{sp,m}(\bm{r}_{\perp})\cdot \bm{E}_{L}(\bm{r}_{\perp},z,t) \mathrm{,}
	\label{eq:envelope_electric_field}
	\end{split}
\end{equation}
which shows that the source current is driven at the laser carrier frequency $\omega_L$.

Let's consider a laser pulse propagating parallel to the interface. For a general cylindrical or planar geometry, write the prescribed electric field as
\begin{equation}
	\bm{E}_L(\bm{r}_{\perp}, z ,t) = \bm{E}_{L0}(\bm{r}_{\perp}, \phi) g(\zeta_L) e^{i(k_L z + m_L \phi - \omega_L t)} + c.c. \mathrm{,} \nonumber
\end{equation}
where $m_L$ is the azimuthal index and $g(\zeta_L)$ the longitudinal pulse envelope and $\zeta_L=z-v_L t$. $\bm{E}_{L0}$ is the complex transverse vector profile, including both spatial dependence and polarisation structure. For an ordinary Gaussian beam without angular momentum, $m_L=0$.
The cylindrical eigenmode can be written as
\begin{equation}
	\bm{E}_{sp,m} = \tilde{\bm{E}}_{sp,m}(r) e^{i m \phi}e^{ik_{sp,m}z - i \omega_{sp,m}t}
\end{equation}
Projecting onto an RSP eigenmode of frequency $\omega_{sp,m}$, axial wavenumber $k_{sp,m}$ and azimuthal index $m$, the driven equation becomes
\begin{equation}
	(\partial_t + v_{g,m} \partial_z + \Gamma_m) A_m = G_m \, g(\zeta_L) e^{i(\Delta k_m z - \Delta \omega_m t)}
\end{equation}
with $\Delta \omega_m = \omega_L - \omega_{sp,m}$ and $\Delta k_m = k_L - k_{sp,m}$. The coupling coefficient for cylindrical geometry is defined as
\begin{equation}
\begin{split}
	G_m = - \frac{i e^2}{4 \omega_L m_e \gamma_q \mathcal{N}_{sp,m}} & \int d \mathbb{A} \, n_0(\bm{r}_{\perp}) \tilde{\bm{E}}_{sp,m}^{*}(\bm{r}_{\perp}) \\
	& \cdot \bm{E}_{L0}(\bm{r}_{\perp}, \phi) e^{i(m_L-m)\phi} \mathrm{.}
	\label{eq:G_m}
	\end{split}
\end{equation}
In the planar geometry, the explicit factor $e^{i(m_L-m)\phi}$ is absent.
It is easy to see that the excitation of RSP directly driven by the electric field has the same model structure as the ponderomotive-driven case, except for an additional fast phase factor $e^{i(\Delta k_m z - \Delta \omega_m t)}$.  Therefore, efficient growth requires both frequency resonance $\omega_L \approx \omega_{sp,m}$ and phase matching $k_L \approx k_{sp,m}$.

The exact solution is then given by
\begin{equation}
\begin{split}
	A_m(z,t) = & e^{i(\Delta k_m z - \Delta \omega_m t)} G_m \int_0^{\infty} ds \\
	& e^{-(\Gamma_m -i \Delta_m) s} g(\zeta_L - \Delta v_m s) \mathrm{,}
\end{split}
\label{eq:envelope_electric_general} 
\end{equation}
where $\Delta_m = \Delta \omega_m - v_{g,m} \Delta k_m$ and $\Delta v_m = v_{g,m} -v_L$. 
For a Gaussian laser pulse $g(\zeta_L) = \exp (-2\zeta_L^2/L_0^2)$, Eq.~\eqref{eq:envelope_electric_general} becomes 
\begin{equation}
	\begin{split}
		A_m(\zeta_L,t) = & \frac{G_m}{2} \sqrt{\frac{\pi}{2}} \frac{L_0}{|\Delta v_m|} e^{i(\Delta k_m z - \Delta \omega_m t)}  \\
		& \exp \left[ \frac{\Gamma_m-i\Delta_m}{\Delta v_m} \left( -\zeta_L + \frac{(\Gamma_m - i \Delta_m)L_0^2}{8 \Delta v_m}\right ) \right]  \\
		& \times \mathrm{erfc}\left( \frac{(\Gamma_m-i \Delta_m)L_0}{2 \sqrt{2} |\Delta v_m|} - \mathrm{sgn} (\Delta v_m) \frac{\sqrt{2} \zeta_L}{L_0}\right ) \mathrm{,}
		\label{eq:electric_full_Am}
	\end{split} 
\end{equation}
which is similar to Eq.~\eqref{eq:A_solve_erfc}. It is easy to see that the direct-field case differs from the ponderomotive one only through the complex detuning $\Delta_m$.

\subsection{Linearly polarised (LP) laser pulse}
For an LP laser pulse, the laser field can be written as
\begin{equation}
	\bm{E}_L^{\rm LP} (r_{\perp}, \phi, z, t) = E_0 F(r_{\perp}) g(\zeta_L) \hat{\bm{e}}_{\rm LP}(\phi)  e^{i(k_L z - \omega_L t)} + c.c. \mathrm{,} \nonumber
\end{equation}
where $\hat{\bm{e}}_{LP}$ is the polarisation vector and $m_L=0$.
The coupling coefficient becomes
\begin{equation}
\begin{split}
	G^{\rm LP}=-\frac{ie^2 E_0}{4 \omega_L \gamma_q m_e \mathcal{N}_{sp}} 
	& \int d\mathbb{A}n_0(\bm{r}_{\perp}) \bm{E}_{sp}^{*} (\bm{r}_{\perp}, \phi) \\
	& \cdot \hat{\bm{e}}_{\rm LP}(\phi) F(\bm{r}_{\perp})\mathrm{,}
\end{split}
\end{equation}
where $F(\bm{r}_{\perp})$ is the transverse amplitude profile. 

\textit{On a planar surface} occupying in $yz$-plane at $x=0$, $\hat{\bm{e}}_{\rm LP}^{(pl)}=e_{Lx} (x)\hat{\bm{x}} + e_{Lz} (x) \hat{\bm{z}}$ with $|e_{Lx}|^2 + |e_{Lz}|^2=1$. By using a planar RSP field $\bm{E}_{sp}=(E_{sp,x}, 0, E_{sp,z})$, the LP coupling coefficient is 
\begin{equation}
\begin{split}
		G_{\rm pl}^{\rm LP} = &  -\frac{ie^2 E_0}{4 \omega_L \gamma_q m_e \mathcal{N}_{sp}} \int^{-\infty}_0 dx \, n_0(x) F(x)   \\
		& \cdot\left[ E^*_{sp,x}(x)e_{Lx}(x) + E^*_{sp,z}(x) e_{Lz}(x)  \right ]    \mathrm{.}
\end{split}
\end{equation}
Since RSP is TM and typically dominated by the normal field near the planar surface, a good approximation is 
\begin{equation}
		G_{\rm pl}^{\rm LP} \approx -\frac{i e^2 E_0 e_{Lx}}{4 \omega_L \gamma_q m_e \mathcal{N}_{sp}}  \int^{-\infty}_0 dx \, n_0(x) F(x) E^*_{sp,x}(x)  \mathrm{,} 
		\label{eq:G_pl_LP}
\end{equation}
which indicates that the direct LP drive is strongest when the laser has a strong normal electric field component at the interface.  
This implies a $p$-selection rule. 
For a single monochromatic plane wave on a perfectly infinite planar interface, the source carries one fixed tangential wavevector $k_{L, \parallel} \leq \omega_L/c$, whereas the RSP eigenmode requires $k_{sp}>\omega_{sp}/c$. Therefore, the projected source is nonzero only if $k_{L, \parallel}=k_{sp}$, which cannot be satisfied for the ordinary planar RSP mode. This explains why a finite tangential spectrum is required. For example, a finite longitudinal extent along the interface, an edge, tip curvature, cylindrical curvature, or a grating. A finite normal spot size alone modifies the overlap but does not by itself supply the missing tangential momentum.  

\textit{On a cylindrical surface}, we consider a laser pulse propagating along $\hat{z}$-direction inside a tube and linearly polarised as $\hat{\bm{e}}_{\rm LP}^{(cy)}=\cos (\phi-\alpha) \, \hat{\bm{r}} - \sin (\phi - \alpha) \, \hat{\bm{\phi}}$ as shown in Fig.~\ref{fig:geometry}(b). The laser field is written as
\begin{equation}
	\bm{E}_{L}^{LP,cy} (r,\phi,z,t) = E_0 F(r) g(\zeta_L) \, \hat{\bm{e}}_{\rm LP}^{(cy)}\, e^{i(k_Lz-\omega_L t)} + c.c. \mathrm{,}\nonumber
\end{equation}
 where the parameter $\alpha$ controls the polarisation direction. For $x$-polarisation, we set $\alpha=0$.
The dominant overlap becomes
\begin{equation}
	G_m^{LP, cy} \propto \int_{0}^{2 \pi} d\phi e^{-im\phi} \cos (\phi-\alpha) = \pi (e^{-i\alpha}\delta_{m,1} + e^{i\alpha}\delta_{m,-1}) \nonumber\mathrm{,}
\end{equation}
which indicates that a transverse LP pulse does not drive the axisymmetric $m=0$. This explains why the ponderomotive drive is dominant even in the high-intensity LP laser-driven case~\cite{Lei:2025aa}. 
Instead, it excites an equal superposition of the two degenerate helical states $m=\pm 1$, equivalently the real $\cos \phi$ surface mode, as the LP electric field itself carries an angular dependence on $\cos \phi$, as shown in Fig.~\ref{fig:sel_rule} (e) and (f).
The explicit form of the coupling coefficient is given as
\begin{equation}
	\begin{split}
		& G_{m}^{LP,cy} = -\frac{ie^2 E_0}{4 \omega_L \gamma_q m_e \mathcal{N}_{sp,m}}  
		\int_R^{\infty} r dr \, n_0(r) F(r) \int_0^{2\pi} d\phi  \\
		& \cdot e^{-im\phi} \left[  \tilde{E}^*_{sp,r,m}(r) \cos(\phi-\alpha) - \tilde{E}^*_{sp,\phi,m}(r) \sin(\phi-\alpha)\right ]
		\label{eq:G_cy_m_LP}
	\end{split}
\end{equation}

\subsection{Circularly polarised (CP) laser pulse}
For a CP laser pulse, 
\begin{equation}
	\bm{E}_{L}^{CP}(r,\phi,z,t) = E_0 F(r) g(\zeta_L) \hat{\bm{e}}_{\rm CP}^{(\sigma)} e^{i(k_L z - \omega_L t)} + c.c. \mathrm{,}
\end{equation}
with $\hat{\bm{e}}_{\rm CP}^{(\sigma)}=(\hat{\bm x} + i \sigma \hat{\bm y})/\sqrt{2}=e^{i\sigma \phi}(\hat{\bm{r}} + i \sigma \hat{\bm{\phi}})/\sqrt{2}$ where $\sigma=\pm 1$ denotes the sign of helicity. Then, the coupling coefficient becomes
\begin{equation}
\begin{split}
		G^{CP} = -\frac{i e^2  E_0}{4 \omega_L \gamma_q m_e \mathcal{N}_{sp}} \int d \mathbb{A} \, n_0(r_{\perp}) F(r_{\perp})\bm{E}_{sp}^*(r_{\perp}) \cdot \hat{\bm{e}}_{\rm CP}^{(\sigma)}  \mathrm{.} 
\end{split}
	\label{eq:G_m_CP}
\end{equation}

\textit{On a planar surface}, relative to the incidence plane, a CP wave decomposes as $\hat{\bm{e}}_{\rm CP}=(\hat{\bm{p}} + i \sigma \hat{\bm{s}})/\sqrt{2}$. As RSP is TM, only the $p$-polarised part couples. So, for equal total field amplitude, the usable coupling of a CP plane wave into a planar TM RSP is reduced by a factor of $1/\sqrt{2}$ relative to a pure $p$-polarised LP wave, while the relativistic detuning is stronger because $\gamma_q^{CP}>\gamma_q^{LP}$. The same translational symmetry prohibition still applies. Therefore, within the present planar TM-coupling model, CP does not provide an intrinsic coupling advantage over a purely p-polarised LP drive.

\textit{On a cylindrical surface}, a laser pulse propagates along the tube axis, $\hat{e}_{\rm CP}=e^{i \sigma \phi} (\hat{\bm{r}} + i \sigma \hat{\bm{\phi}})/\sqrt{2}$. 
For the TM-dominant or strongly surface-confined part of the cylindrical RSP, the dominant electric overlap is usually through the radial component, for which $\hat{\bm{e}}_{\rm CP} \cdot \hat{\bm{r}}= e^{i \sigma \phi}/\sqrt{2}$. As $\int_0^{2\pi} d \phi e^{-i m \phi} e^{i\sigma \phi} = 2\pi \delta_{m,\sigma}$, the explicit cylindrical coupling coefficient is 
\begin{equation}
\begin{split}
	G_{m}^{CP,cy} = & - \frac{i \pi e^2 E_0}{2 \sqrt{2} \omega_L \gamma_q m_e \mathcal{N}_{sp,m}} 
	 \delta_{m,\sigma} \int_R^{\infty} r dr \, n_0(r) F(r) \\
	&  \quad \quad \cdot \left [ \tilde{E}^*_{sp, r, \sigma} (r) +  i \sigma \tilde{E}^*_{sp, \phi, \sigma} (r)\right ]  \mathrm{,}  \label{eq:G_cy_m_CP}
\end{split}
\end{equation}
which indicates that a right or left CP pulse selectively excites a single helical mode $m=\pm 1$, as shown in Fig.~\ref{fig:sel_rule} (g)-(i) for $\sigma=+1$ and (j)-(l) for $\sigma=-1$.
As a result, the polarisation becomes a mode selector on a cylindrical surface. 

\begin{figure}
	\includegraphics[width=0.5\textwidth]{fig6_selection_rule.jpg}
	\caption{Selection rule: $E_z$ component of RSP field inside the vacuum cylindrical channel driven by ponderomotive (a-c), LP electric drive (d-f) and CP  electric drive for $\sigma=+1$ (g-i) and $\sigma=-1$ (j-l). The first column shows the amplitude $|E_z|$. The second and third columns show $\Re(E_z)$ at phase $0$ and $0.25 \pi$, respectively. }
	\label{fig:sel_rule}
\end{figure}

\section{Discussion and conclusion}
In this paper, we present a classical theory of RSP excitation from Maxwell's equations and fluid equations. 
The theoretical framework presented here is general. The physical RSP field is determined by the overlap efficiency between the eigenfields and the external drive, while the eigenfields are intrinsic to the surface geometry.
Two different surface geometries and drives are analytically studied. Before including changes of eigenmode, density evacuation, surface softening, and detuning, an axisymmetric ponderomotive force can drive an axisymmetric mode $m=0$ that can generate a high gradient plasma wakefield. The explicit prefactor scales as $|S_0|\propto a_0^2/\gamma_q^2$, which favours stronger mode driving at fixed mode normalisation. 
For $a_0\gg 1$,$\gamma_q\sim a_0 \sqrt{s_p}$, so the explicit source factor saturates.
 For $a_0 \ll 1$, $\gamma_q\sim 1$, and the familiar nonrelativistic scaling $|S_0|\propto a_0^2$ is recovered.
The direct laser electric field can drive non-axisymmetric cylindrical modes according to the azimuthal selection rule. This can enable controlled RSP generation by the incident laser field, thereby supporting efficient electron acceleration and manipulation. While the explicit direct-field prefactor scales as $|G_m|\propto a_0/\gamma_q$, the excited RSP field imprints the characteristics of the drive through the selection rule. This provides a controllable way to manipulate the electron dynamics precisely.
Compared to a planar surface, a cylindrical surface can provide advantages in the accelerating field generation and mode control.

The external drive can also be applied to different types of sources, each of which should provide distinctive advantages. For example, a beat-wave drive is possible to enable a resonant excitation~\cite{Lei:2026aa}. 
 With a charged electron beam of uniform density, the external drive is $f_{\rm ext} = 2 \pi e^2 n_b r_b^2/ \gamma_b^2 \bm{r}$, where $n_b$, $r_b$ and $\gamma_b$ are beam number density, radius and averaged Lorentz factor, respectively. In this case, the ultrahigh gradient nonlinear plasma wakefield can be generated due to the absence of carrier frequency effects~\cite{Lei:2025ab}.
The RSP excitation with different drives can be solved similarly by using the methods presented in this paper.

The coupling mechanism in RSP excitation presents the selection rules as shown in Fig.~\ref{fig:sel_rule}. For example, on a cylindrical surface, an axisymmetric ponderomotive source leads to the $m=0$ mode. A CP laser pulse with helicity $\sigma=\pm 1$ selectively excites the single helical mode $m=\sigma$. An LP laser excites the $m=\pm 1$ subspace as an equal superposition, which gives a real dipole-like surface pattern.
In principle, Eq.~\eqref{eq:G_m} implies a more general mode selection rule by the factor $e^{i(m_L-m)}\phi$. For example, with an LG laser,  the high-order RSP mode can be selected accordingly~\cite{Jin:2023aa}. 
The azimuthal selection rules for non-axisymmetric modes remain valid even when Eq.~\eqref{eq:disp_cyl} is replaced by the full hybrid cylindrical dispersion relation because they follow from angular orthogonality.
This feature provides a flexible way to control the RSP excitation.

In this work, the relativistic correction is treated by assuming the frozen laser envelope during interaction. Specifically, the local laser quiver factor $\gamma_L(\bm{r}_{\perp}, z,t)=\sqrt{1+s_p a_L(\bm{r}_{\perp}, z, t)^2}$ is replaced by a representative surface value $\gamma_q=\gamma_L|_{\rm surf}=\sqrt{1+s_p a_s^2}$, where $a_s$ denotes the representative normalised laser amplitude at the surface, or more generally in the dominant RSP-overlap region.
This single value of $\gamma_q$ is then used in the relativistic Drude dielectric response, in the source-current response, in the RSP dispersion relation, and in the mode normalisation. As a result, the eigenmode problem remains stationary, and the driven amplitude equation can be solved analytically. 
This approximation is valid when the laser envelope changes slowly compared with the RSP oscillation and when the RSP samples only a region where the laser intensity is approximately constant. This requires, approximately, $|\partial_t \gamma_L/\gamma_L|\ll \omega_{sp}$, $|\partial_z \gamma_L/\gamma_L|\ll k_{sp}$ and $|\partial_{\perp} \gamma_L/\gamma_L|\ll \kappa_2$.
For a Gaussian laser pulse, these conditions correspond roughly to $L_0 \gg \lambda_{sp}$ and $w_0 \gg \kappa_2^{-1}$.
Therefore, the approximation is most appropriate for long pulses, broad laser spots, moderate relativistic strengths, and relatively sharp, stable surfaces. It is also useful to obtain analytic scaling laws, dispersion shifts, and selection rules rather than a fully time-dependent nonlinear description.

The approximation becomes inaccurate when the laser envelope varies strongly during the interaction. This can occur for ultrashort pulses with $L_0\sim \lambda_{sp}$, tightly focused beams with $w_0 \sim \kappa_2^{-1}$, or strongly relativistic pulses for which $\gamma_L$ changes significantly across the RSP skin layer. It is also insufficient when the laser front causes density evacuation, surface expansion, surface softening, preplasma formation, or relativistic transparency. In these regimes, the local plasma response cannot be represented by a single constant $\gamma_q$, and the source coefficient is no longer a constant but a space-time-dependent function. As a result, the closed-form error-function solution in Eq.~\eqref{eq:A_solve_erfc} and \eqref{eq:electric_full_Am} should be replaced by the Green-function integral.

The present theory can therefore be extended in several systematic ways. The simplest extension is a source-level non-frozen model, in which the eigenmode quantities $\bm{E}_{sp}$, $\mathcal{N}_{sp}$, $v_g$ and $\omega_{sp}$ are still evaluated using a representative $\gamma_q$, but the source terms retain the local factor $\gamma_L(\bm{r}_{\perp}, z, t)$. A more complete extension is an adiabatic time-dependent eigenmode theory, in which the dielectric function becomes
\begin{equation}
	\varepsilon_2(\bm{r}_{\perp}, z, t; \omega) = 1 - \frac{\omega_p^2(\bm{r}_{\perp}, z, t)}{\gamma_L(\bm{r}_{\perp}, z, t) \omega^2} \mathrm{,}
\end{equation}
so that the eigenfield, dispersion relation, mode normalisation, group velocity, and detuning all evolve with the laser envelope and plasma density. In this case, additional terms proportional to $\partial_t \gamma_L$, $\nabla \gamma_L$ and $\partial_t n_0$ should appear in the amplitude equation. Such an extension would allow the theory to describe strong surface deformation, density-gradient formation, relativistic transparency, and strongly nonlinear RSP excitation more quantitatively.

\section{List of Symbols}
\begin{tabularx}{\linewidth}{@{} >{\raggedright\arraybackslash}p{0.23\linewidth}  X @{}}
\toprule
$\omega_{\rm sp},\,k_{\rm sp}$ & RSP eigenfrequency and axial wavenumber \\
$\omega_L,\,k_L$ & Laser carrier frequency and axial wavenumber \\
$\omega_d$ & Driving spectral component used in projection \\
$q_\parallel$ & Planar in-plane Fourier variable (general) \\
$\gamma$ & Relativistic Lorentz factor of the fluid velocity $\bm v$ \\
$\gamma_L(\bm r,t)$ & Local laser quiver factor \\
$\gamma_q$ & Representative quiver factor used in $\varepsilon(\omega)$ \\
$a_0$ & Laser normalised peak amplitude\\
$a_s$ & Laser normalised amplitude at surface \\
$\kappa_1,\,\kappa_2$ & Transverse evanescence constants in vacuum/plasma \\
$\mathcal{N}_{\rm sp}$ & Mode-energy normalisation \\
$v_g$ & Group velocity $(\partial\omega/\partial k_z)_{\rm sp}$ \\
$\nu$ & Collision rate in Drude model  \\
$\Gamma$ & effective modal damping rate \\
\bottomrule
\end{tabularx}

\begin{acknowledgements}
This work has been supported by the Science and Technology Facilities Council (STFC) through the Cockcroft Institute core grant UKRI1887.
Guoxing Xia acknowledges the support from the Cockcroft Institute Core Grant No. ST/V001612/1.
\end{acknowledgements}

\bibliography{rsp.bib}

@article{Lei:2025oty,
	author = {Lei, Bifeng and Zhang, Hao and Seipt, Daniel and Bonatto, Alexandre and Qiao, Bin and Resta-L{\'o}pez, Javier and Xia, Guoxing and Welsch, Carsten},
	date = {2025/11/12/},
	day = {12},
	doi = {10.1103/cnym-16hc},
	id = {10.1103/cnym-16hc},
	j1 = {PRL},
	journal = {Physical Review Letters},
	journal1 = {Phys. Rev. Lett.},
	month = {11},
	number = {20},
	pages = {205001--},
	publisher = {American Physical Society},
	title = {Coherent Synchrotron Radiation by Excitation of Surface Plasmon Polariton on Near-Critical Solid Microtube Surface},
	url = {https://link.aps.org/doi/10.1103/cnym-16hc},
	volume = {135},
	year = {2025}}

@inproceedings{Hotta:2004aa,
	author = {Hotta, M. and Hano, M. and Awai, I.},
	booktitle = {34th European Microwave Conference, 2004.},
	pages = {439-442},
	title = {Surface waves along a boundary of single-negative and double-positive materials},
	volume = {1},
	year = {2004}}

@article{Macedo:2019aa,
	author = {Mac{\^e}do, Rair and Camley, Robert E.},
	date = {2019/01/29/},
	day = {29},
	doi = {10.1103/PhysRevB.99.014437},
	id = {10.1103/PhysRevB.99.014437},
	j1 = {PRB},
	journal = {Physical Review B},
	journal1 = {Phys. Rev. B},
	month = {01},
	number = {1},
	pages = {014437--},
	publisher = {American Physical Society},
	title = {Engineering terahertz surface magnon-polaritons in hyperbolic antiferromagnets},
	url = {https://link.aps.org/doi/10.1103/PhysRevB.99.014437},
	volume = {99},
	year = {2019}}

@article{Huang:2004aa,
	author = {Huang, Kerwyn Casey and Povinelli, M. L. and Joannopoulos, John D.},
	doi = {10.1063/1.1775291},
	isbn = {0003-6951},
	journal = {Applied Physics Letters},
	journal1 = {Appl. Phys. Lett.},
	month = {4/27/2026},
	number = {4},
	pages = {543--545},
	title = {Negative effective permeability in polaritonic photonic crystals},
	url = {https://doi.org/10.1063/1.1775291},
	volume = {85},
	year = {2004},
	year1 = {2004/07/26}}

@article{Vinogradov:2006aa,
	author = {Vinogradov, A. P. and Dorofeenko, A. V. and Erokhin, S. G. and Inoue, M. and Lisyansky, A. A. and Merzlikin, A. M. and Granovsky, A. B.},
	date = {2006/07/31/},
	day = {31},
	doi = {10.1103/PhysRevB.74.045128},
	id = {10.1103/PhysRevB.74.045128},
	j1 = {PRB},
	journal = {Physical Review B},
	journal1 = {Phys. Rev. B},
	month = {07},
	number = {4},
	pages = {045128--},
	publisher = {American Physical Society},
	title = {Surface state peculiarities in one-dimensional photonic crystal interfaces},
	url = {https://link.aps.org/doi/10.1103/PhysRevB.74.045128},
	volume = {74},
	year = {2006}}

@article{Ali:2018aa,
	author = {Ali, Rashid and Zamir, Burhan and Shah, H. A.},
	date = {2018/03/01/},
	doi = {https://doi.org/10.1016/j.rinp.2017.11.037},
	isbn = {2211-3797},
	journal = {Results in Physics},
	pages = {243--248},
	title = {Transverse electric surface waves in a plasma medium bounded by magnetic materials},
	url = {https://www.sciencedirect.com/science/article/pii/S2211379717315905},
	volume = {8},
	year = {2018}}

@article{Smith:2004aa,
	author = {Smith, D. R. and Pendry, J. B. and Wiltshire, M. C. K.},
	date = {2004/08/06},
	doi = {10.1126/science.1096796},
	journal = {Science},
	journal1 = {Science},
	journal2 = {Science},
	month = {2026/04/27},
	n2 = {Recently, artificially constructed metamaterials have become of considerable interest, because these materials can exhibit electromagnetic characteristics unlike those of any conventional materials. Artificial magnetism and negative refractive index are two specific types of behavior that have been demonstrated over the past few years, illustrating the new physics and new applications possible when we expand our view as to what constitutes a material. In this review, we describe recent advances in metamaterials research and discuss the potential that these materials may hold for realizing new and seemingly exotic electromagnetic phenomena.},
	number = {5685},
	pages = {788--792},
	publisher = {American Association for the Advancement of Science},
	title = {Metamaterials and Negative Refractive Index},
	type = {doi: 10.1126/science.1096796},
	url = {https://doi.org/10.1126/science.1096796},
	volume = {305},
	year = {2004},
	year1 = {2004}}

@article{Little:2012aa,
	author = {Little, C. E. and Anufriev, R. and Iorsh, I. and Kaliteevski, M. A. and Abram, R. A. and Brand, S.},
	date = {2012/12/14/},
	day = {14},
	doi = {10.1103/PhysRevB.86.235425},
	id = {10.1103/PhysRevB.86.235425},
	j1 = {PRB},
	journal = {Physical Review B},
	journal1 = {Phys. Rev. B},
	month = {12},
	number = {23},
	pages = {235425--},
	publisher = {American Physical Society},
	title = {Tamm plasmon polaritons in multilayered cylindrical structures},
	url = {https://link.aps.org/doi/10.1103/PhysRevB.86.235425},
	volume = {86},
	year = {2012}}

@article{Shen:2024aa,
	author = {Shen, Xiaofei and Pukhov, Alexander and Qiao, Bin},
	date = {2024/03/07},
	doi = {10.1038/s42005-024-01575-z},
	id = {Shen2024},
	isbn = {2399-3650},
	journal = {Communications Physics},
	number = {1},
	pages = {84},
	title = {High-flux bright x-ray source from femtosecond laser-irradiated microtapes},
	url = {https://doi.org/10.1038/s42005-024-01575-z},
	volume = {7},
	year = {2024}}

@software{vay_2025_17261711,
	author = {Vay, Jean-Luc and Almgren, Ann and Amorim, L{\'\i}gia Diana and Andriyash, Igor and Angus, Justin Ray and Belkin, Daniel and Bizzozero, David and Blelly, Aurore and Clark, Stephen Eric and Fedeli, Luca and Formenti, Arianna and Garten, Marco and Ge, Lixin and Gott, Kevin and Harrison, Cyrus and Huebl, Axel and Giacomel, Lorenzo and Groenewald, Roelof E. and Grote, David and Gu, Junmin and Haseeb, Muhammad and Jambunathan, Revathi and Klion, Hannah and Kumar, Prabhat and Th{\'e}venet, Maxence and Richardson, Glenn and Shapoval, Olga and Lehe, Remi and Loring, Burlen and Miller, Phil and Myers, Andrew and Rheaume, Elisa and Rowan, Michael E. and Sandberg, Ryan Thor and Scherpelz, Peter and Weichman, Kale and Yang, Eloise and Zaim, Ne{\"\i}l and Zhang, Weiqun and Zhao, Yinjian and Zhu, Kevin Z. and Zoni, Edoardo},
	doi = {10.5281/zenodo.17261711},
	month = oct,
	publisher = {Zenodo},
	swhid = {swh:1:dir:67ea49560ec391114918a3ab29effb069e9b93da ;origin=https://doi.org/10.5281/zenodo.4571577;vis it=swh:1:snp:f87f5afe9298f76157699bbe680c60bd620de 86d;anchor=swh:1:rel:c55f8c0e072d4a6d0a72dfc6ed57f ed7f7374650;path=BLAST-WarpX-warpx-46e6b71},
	title = {BLAST-WarpX/warpx: 25.10},
	url = {https://doi.org/10.5281/zenodo.17261711},
	version = {25.10},
	year = 2025}

@article{Stegeman:1983aa,
	author = {Stegeman, G. I. and Wallis, R. F. and Maradudin, A. A.},
	date = {1983/07/01},
	doi = {10.1364/OL.8.000386},
	j2 = {Opt. Lett.},
	journal = {Optics Letters},
	journal1 = {Optics Letters},
	journal2 = {Opt. Lett.},
	journal3 = {Opt. Lett.},
	number = {7},
	pages = {386--388},
	publisher = {Optica Publishing Group},
	title = {Excitation of surface polaritons by end-fire coupling},
	url = {https://opg.optica.org/ol/abstract.cfm?URI=ol-8-7-386},
	volume = {8},
	year = {1983}}

@inbook{Greffet:2012aa,
	address = {Berlin, Heidelberg},
	author = {Greffet, Jean-Jacques},
	booktitle = {Plasmonics: From Basics to Advanced Topics},
	date = {2012//},
	doi = {10.1007/978-3-642-28079-5{\_}4},
	editor = {Enoch, Stefan and Bonod, Nicolas},
	id = {Greffet2012},
	isbn = {978-3-642-28079-5},
	pages = {105--148},
	publisher = {Springer Berlin Heidelberg},
	title = {Introduction to Surface Plasmon Theory},
	url = {https://doi.org/10.1007/978-3-642-28079-5_4},
	year = {2012}}

@article{Yu:2019aa,
	author = {Yu, Huakang and Peng, Yusi and Yang, Yong and Li, Zhi-Yuan},
	date = {2019/04/11},
	doi = {10.1038/s41524-019-0184-1},
	id = {Yu2019},
	isbn = {2057-3960},
	journal = {npj Computational Materials},
	number = {1},
	pages = {45},
	title = {Plasmon-enhanced light--matter interactions and applications},
	url = {https://doi.org/10.1038/s41524-019-0184-1},
	volume = {5},
	year = {2019}}

@inbook{Lei:2026intch,
	address = {London},
	author = {Resta-L{\'o}pez, Javier and Bonatto, Alexandre and Xia, Guoxing and Lei, Bifeng and Welsch, Carsten},
	booktitle = {Plasma Science - Current Developments, Applications, and Future Directions},
	cn = {11},
	doi = {10.5772/intechopen.1013683},
	editor = {Resta-L{\'o}pez, Javier and Bonatto, Alexandre},
	isbn = {978-1-83635-607-3},
	month = {2026-03-26},
	n2 = {Surface plasmons (SPs) are collective oscillations of free electrons that are confined to the interface between conducting (e.g., plasma) and dielectric (e.g., vacuum) materials. When SPs couple strongly with electromagnetic waves (photons), they form hybrid quasiparticles called surface plasmon polaritons (SPPs). This coherent behaviour can be driven by external beams such as photons or charged particle beams, through the long-range Coulomb interactions between conduction electrons. When external sources are highly intense, with strong fields that can quickly accelerate free electrons to relativistic speeds, the dynamics of SPs differ from those in the low-energy case. In this chapter, we demonstrate that the SPs driven by high-intensity laser or particle beams can provide TV/m-level electromagnetic fields. Charged particles can be efficiently accelerated if they are properly placed and maintained in these fields. Three different mechanisms of SP-based particle acceleration are then proposed by using particle-in-cell simulations, including leaky and bubble wakefield acceleration and direct electron acceleration due to the resonant coupling of SPs. These results demonstrate that the efficient excitation of SPs in strong fields could provide a promising solution for the next generation of ultra-compact particle accelerators. It can offer unprecedented benefits in various advanced applications due to its ultra-high acceleration gradient and flexibility.},
	publisher = {IntechOpen},
	title = {Surface Plasmons in Strong Fields and Their Applications to Beam Acceleration},
	url = {https://doi.org/10.5772/intechopen.1013683},
	year = {2026-01-27},
	year1 = {2026}}

@article{Stern:1960aa,
	author = {Stern, E. A. and Ferrell, R. A.},
	date = {1960/10/01/},
	day = {01},
	doi = {10.1103/PhysRev.120.130},
	id = {10.1103/PhysRev.120.130},
	j1 = {PR},
	journal = {Physical Review},
	journal1 = {Phys. Rev.},
	month = {10},
	number = {1},
	pages = {130--136},
	publisher = {American Physical Society},
	title = {Surface Plasma Oscillations of a Degenerate Electron Gas},
	url = {https://link.aps.org/doi/10.1103/PhysRev.120.130},
	volume = {120},
	year = {1960}}

@article{Lei:2026aa,
	author = {Lei, Bifeng and Zhang, Hao and Bonatto, Alexandre and Liu, Bin and Resta-Lopez, Javier and Zepf, Matt and Xia, Guoxing and Welsch, Carsten},
	journal = {arXiv preprint arXiv:2602.12789},
	title = {Resonant Excitation of Surface Plasmon for Wakefield Acceleration by Beating GW Lasers on Smooth Cylindrical Surface},
	year = {2026}}

@book{Bohren:2008aa,
	author = {Bohren, C.F. and Huffman, D.R. %@ 9783527618163},
	date = {2008},
	publisher = {Wiley},
	title = {Absorption and Scattering of Light by Small Particles},
	url = {https://books.google.co.uk/books?id=ib3EMXXIRXUC},
	year = {2008}}

@book{agranovich2012surface,
	author = {Agranovich, V.M.},
	isbn = {9780444598691},
	publisher = {North Holland},
	series = {Modern Problems in Condensed Matter Sciences},
	title = {Surface Polaritons},
	url = {https://books.google.co.uk/books?id=wKJpV8c2ZtQC},
	year = {2012}}

@article{Zayats:2005aa,
	author = {Zayats, Anatoly V. and Smolyaninov, Igor I. and Maradudin, Alexei A.},
	date = {2005/03/01/},
	doi = {https://doi.org/10.1016/j.physrep.2004.11.001},
	isbn = {0370-1573},
	journal = {Physics Reports},
	number = {3},
	pages = {131--314},
	title = {Nano-optics of surface plasmon polaritons},
	url = {https://www.sciencedirect.com/science/article/pii/S0370157304004600},
	volume = {408},
	year = {2005}}

@article{McCay:2025aa,
	author = {McCay, A. and McIlvenny, A. and Macchi, A. and Romagnani, L. and Martin, P. and Cavanagh, O. and Molloy, D. P. and Lancia, L. and Dzelzainis, T. and Ahmed, H. and Sarma, J. and Kar, S. and Margarone, D. and Borghesi, M.},
	date = {2025/10/01/},
	day = {01},
	doi = {10.1103/y29y-f63h},
	id = {10.1103/y29y-f63h},
	j1 = {PRL},
	journal = {Physical Review Letters},
	journal1 = {Phys. Rev. Lett.},
	month = {10},
	number = {14},
	pages = {145001--},
	publisher = {American Physical Society},
	title = {Surface Wave Electron Acceleration from Flat Foils at Parallel Laser Incidence},
	url = {https://link.aps.org/doi/10.1103/y29y-f63h},
	volume = {135},
	year = {2025}}

@article{Willingale:2011aa,
	author = {Willingale, L. and Nilson, P. M. and Thomas, A. G. R. and Cobble, J. and Craxton, R. S. and Maksimchuk, A. and Norreys, P. A. and Sangster, T. C. and Scott, R. H. H. and Stoeckl, C. and Zulick, C. and Krushelnick, K.},
	date = {2011/03/10/},
	day = {10},
	doi = {10.1103/PhysRevLett.106.105002},
	id = {10.1103/PhysRevLett.106.105002},
	j1 = {PRL},
	journal = {Physical Review Letters},
	journal1 = {Phys. Rev. Lett.},
	month = {03},
	number = {10},
	pages = {105002--},
	publisher = {American Physical Society},
	title = {High-Power, Kilojoule Class Laser Channeling in Millimeter-Scale Underdense Plasma},
	url = {https://link.aps.org/doi/10.1103/PhysRevLett.106.105002},
	volume = {106},
	year = {2011}}

@article{Willingale:2013aa,
	author = {Willingale, L and Thomas, A G R and Nilson, P M and Chen, H and Cobble, J and Craxton, R S and Maksimchuk, A and Norreys, P A and Sangster, T C and Scott, R H H and Stoeckl, C and Zulick, C and Krushelnick, K},
	date = {2013/02/18},
	doi = {10.1088/1367-2630/15/2/025023},
	isbn = {1367-2630;},
	journal = {New Journal of Physics},
	number = {2},
	pages = {025023},
	publisher = {IOP Publishing},
	title = {Surface waves and electron acceleration from high-power, kilojoule-class laser interactions with underdense plasma},
	url = {https://dx.doi.org/10.1088/1367-2630/15/2/025023},
	volume = {15},
	year = {2013}}

@article{Riconda:2015aa,
	author = {Riconda, C. and Raynaud, M. and Vialis, T. and Grech, M.},
	doi = {10.1063/1.4923443},
	isbn = {1070-664X},
	journal = {Physics of Plasmas},
	journal1 = {Phys. Plasmas},
	month = {8/8/2025},
	number = {7},
	pages = {073103},
	title = {Simple scalings for various regimes of electron acceleration in surface plasma waves},
	url = {https://doi.org/10.1063/1.4923443},
	volume = {22},
	year = {2015},
	year1 = {2015/07/10}}

@article{Zaidi:1991aa,
	author = {Zaidi, Saleem H. and Yousaf, M. and Brueck, S. R. J.},
	date = {1991/04/01},
	doi = {10.1364/JOSAB.8.000770},
	j2 = {J. Opt. Soc. Am. B},
	journal = {Journal of the Optical Society of America B},
	journal1 = {Journal of the Optical Society of America B},
	journal2 = {J. Opt. Soc. Am. B},
	journal3 = {J. Opt. Soc. Am. B},
	number = {4},
	pages = {770--779},
	publisher = {Optica Publishing Group},
	title = {Grating coupling to surface plasma waves. I. First-order coupling},
	url = {https://opg.optica.org/josab/abstract.cfm?URI=josab-8-4-770},
	volume = {8},
	year = {1991}}

@article{Jin:2023aa,
	author = {Jin, S and Yao, Y L and Lei, B F and Chen, G Y and Zhou, C T and Zhu, S P and He, X T and Qiao, B},
	date = {2023/09/01/},
	doi = {10.1088/1367-2630/acf683},
	dp = {DOI.org (Crossref)},
	isbn = {1367-2630},
	j2 = {New J. Phys.},
	journal = {New Journal of Physics},
	la = {en},
	month = {2023/10/05/08:01:13},
	number = {9},
	pages = {093030},
	title = {High-energy quasi-monoenergetic proton beam from micro-tube targets driven by Laguerre--Gaussian lasers},
	url = {https://iopscience.iop.org/article/10.1088/1367-2630/acf683},
	volume = {25},
	year = {2023}}

@article{Raynaud:2018aa,
	author = {Raynaud, M and H{\'e}ron, A and Adam, J-C},
	date = {2017/10/31},
	doi = {10.1088/1361-6587/aa8f13},
	isbn = {0741-3335},
	journal = {Plasma Physics and Controlled Fusion},
	number = {1},
	pages = {014021},
	publisher = {IOP Publishing},
	title = {High intensity surface plasma waves, theory and PIC simulations},
	url = {https://dx.doi.org/10.1088/1361-6587/aa8f13},
	volume = {60},
	year = {2018}}

@article{Marini:2021aa,
	author = {Marini, S. and Kleij, P. S. and Amiranoff, F. and Grech, M. and Riconda, C. and Raynaud, M.},
	doi = {10.1063/5.0052599},
	isbn = {1070-664X},
	journal = {Physics of Plasmas},
	journal1 = {Phys. Plasmas},
	month = {8/7/2025},
	number = {7},
	pages = {073104},
	title = {Key parameters for surface plasma wave excitation in the ultra-high intensity regime},
	url = {https://doi.org/10.1063/5.0052599},
	volume = {28},
	year = {2021},
	year1 = {2021/07/19}}

@article{Lei:2025ab,
	author = {Lei, Bifeng and Zhang, Hao and Bontoiu, Cristian and Bonatto, Alexandre and Resta Lopez, Javier and Xia, Guoxing and Qiao, Bin and Welsch, Carsten P.},
	isbn = {1367-2630},
	journal = {New Journal of Physics},
	n2 = {Solid-state materials, such as carbon nanotubes (CNTs), have the potential to support ultra-high accelerating fields in the TV/m range for charged particle acceleration. In this study, we explore the feasibility of using nanostructured CNTs forest to develop plasma-based accelerators at the 100sTeV-level, driven by high-density, ultra-relativistic electron beams, using fully three-dimensional particle-in-cell simulations. Two different acceleration mechanisms are proposed and investigated: the surface plasmon leakage field and the bubble wakefield. The leakage field, driven by a relatively low-density beam, can achieve an acceleration field up to TV/m, capable of accelerating both electron and positron beams. In particular, due to the direct acceleration by the driving beam, the positron acceleration is highly efficient with an average acceleration gradient of 2.3TeV/m. In contrast, the bubble wakefield mechanism allows significantly higher acceleration fields, e.g. beyond 400TV/m, with a much higher energy transfer efficiency of 66.7{\%}. In principle, electrons can be accelerated to PeV energies over distances of several meters. If the beam density is sufficiently high, the CNT target will be completely blown out, where no accelerating field is generated. Its threshold has been estimated. Two major challenges in these schemes are recognised and investigated. Leveraging the ultra-high energy and charge pumping rate of the driving beam, the nanostructured CNTs also offer significant potential for a wide range of advanced applications. This work represents a promising avenue for the development of ultra-compact, high-energy particle accelerators. We also outline conceptual experiments using currently available facilities, demonstrating that this approach is experimentally accessible.},
	title = {100s TeV/m-Level Particle Accelerators Driven by High-density Electron Beams in Micro Structured Carbon Nanotube Forest Channel},
	url = {http://iopscience.iop.org/article/10.1088/1367-2630/adf87d},
	year = {2025}}

@article{Lei:2025aa,
	author = {Lei, Bifeng and Zhang, Hao and Bon{\c t}oiu, Cristian and Bonatto, Alexandre and Mart{\'\i}n-Luna, Pablo and Liu, Bin and Resta L{\'o}pez, Javier and Xia, Guoxing and Welsch, Carsten},
	date = {2025/06/12},
	doi = {10.1088/1361-6587/ade00a},
	isbn = {0741-3335},
	journal = {Plasma Physics and Controlled Fusion},
	number = {6},
	pages = {065036},
	publisher = {IOP Publishing},
	title = {Leaky surface plasmon-based wakefield acceleration in nanostructured carbon nanotubes},
	url = {https://dx.doi.org/10.1088/1361-6587/ade00a},
	volume = {67},
	year = {2025}}

@article{Bonatto:2023aa,
	author = {Bonatto, A. and Xia, G. and Apsimon, O. and Bontoiu, C. and Kukstas, E. and Rodin, V. and Yadav, M. and Welsch, C. P. and Resta-L{\'o}pez, J.},
	doi = {10.1063/5.0134960},
	isbn = {1070-664X},
	journal = {Physics of Plasmas},
	journal1 = {Phys. Plasmas},
	month = {8/7/2025},
	number = {3},
	pages = {033105},
	title = {Exploring ultra-high-intensity wakefields in carbon nanotube arrays: An effective plasma-density approach},
	url = {https://doi.org/10.1063/5.0134960},
	volume = {30},
	year = {2023},
	year1 = {2023/03/08}}

@article{Sarma:2022aa,
	author = {Sarma, J and McIlvenny, A and Das, N and Borghesi, M and Macchi, A},
	date = {2022/07/19},
	doi = {10.1088/1367-2630/ac7d6e},
	isbn = {1367-2630;},
	journal = {New Journal of Physics},
	number = {7},
	pages = {073023},
	publisher = {IOP Publishing},
	title = {Surface plasmon-driven electron and proton acceleration without grating coupling},
	url = {https://dx.doi.org/10.1088/1367-2630/ac7d6e},
	volume = {24},
	year = {2022}}

@article{Shen:2021aa,
	author = {Shen, X. F. and Pukhov, A. and Qiao, B.},
	date = {2021/10/04/},
	day = {04},
	doi = {10.1103/PhysRevX.11.041002},
	id = {10.1103/PhysRevX.11.041002},
	j1 = {PRX},
	journal = {Physical Review X},
	journal1 = {Phys. Rev. X},
	month = {10},
	number = {4},
	pages = {041002--},
	publisher = {American Physical Society},
	title = {Monoenergetic High-Energy Ion Source via Femtosecond Laser Interacting with a Microtape},
	url = {https://link.aps.org/doi/10.1103/PhysRevX.11.041002},
	volume = {11},
	year = {2021}}

@article{Lad:2022aa,
	author = {Lad, Amit D. and Mishima, Y. and Singh, Prashant Kumar and Li, Boyuan and Adak, Amitava and Chatterjee, Gourab and Brijesh, P. and Dalui, Malay and Inoue, M. and Jha, J. and Tata, Sheroy and Trivikram, M. and Krishnamurthy, M. and Chen, Min and Sheng, Z. M. and Tanaka, K. A. and Kumar, G. Ravindra and Habara, H.},
	date = {2022/10/07},
	doi = {10.1038/s41598-022-21210-7},
	id = {Lad2022},
	isbn = {2045-2322},
	journal = {Scientific Reports},
	number = {1},
	pages = {16818},
	title = {Luminous, relativistic, directional electron bunches from an intense laser driven grating plasma},
	url = {https://doi.org/10.1038/s41598-022-21210-7},
	volume = {12},
	year = {2022}}

@article{Zhu:2020aa,
	author = {Zhu, X. M. and Prasad, R. and Swantusch, M. and Aurand, B. and Andreev, A. A. and Willi, O. and Cerchez, M.},
	c7 = {e15},
	db = {Cambridge Core},
	doi = {DOI: 10.1017/hpl.2020.14},
	dp = {Cambridge University Press},
	edition = {2020/04/29},
	isbn = {2095-4719},
	journal = {High Power Laser Science and Engineering},
	pages = {e15},
	publisher = {Cambridge University Press},
	title = {Relativistic electron acceleration by surface plasma waves excited with high intensity laser pulses},
	url = {https://www.cambridge.org/core/product/BA97CBAA5B4CE5B1A0E547E804AE161D},
	volume = {8},
	year = {2020}}

@article{Fedeli:2016aa,
	author = {Fedeli, L. and Sgattoni, A. and Cantono, G. and Garzella, D. and R{\'e}au, F. and Prencipe, I. and Passoni, M. and Raynaud, M. and Kv{\v e}to{\v n}, M. and Proska, J. and Macchi, A. and Ceccotti, T.},
	date = {2016/01/07/},
	day = {07},
	doi = {10.1103/PhysRevLett.116.015001},
	id = {10.1103/PhysRevLett.116.015001},
	j1 = {PRL},
	journal = {Physical Review Letters},
	journal1 = {Phys. Rev. Lett.},
	month = {01},
	number = {1},
	pages = {015001--},
	publisher = {American Physical Society},
	title = {Electron Acceleration by Relativistic Surface Plasmons in Laser-Grating Interaction},
	url = {https://link.aps.org/doi/10.1103/PhysRevLett.116.015001},
	volume = {116},
	year = {2016}}

@article{Brugge:2012aa,
	author = {an der Br{\"u}gge, Daniel and Kumar, Naveen and Pukhov, Alexander and R{\"o}del, Christian},
	date = {2012/03/19/},
	day = {19},
	doi = {10.1103/PhysRevLett.108.125002},
	id = {10.1103/PhysRevLett.108.125002},
	j1 = {PRL},
	journal = {Physical Review Letters},
	journal1 = {Phys. Rev. Lett.},
	month = {03},
	number = {12},
	pages = {125002--},
	publisher = {American Physical Society},
	title = {Influence of Surface Waves on Plasma High-Order Harmonic Generation},
	url = {https://link.aps.org/doi/10.1103/PhysRevLett.108.125002},
	volume = {108},
	year = {2012}}

@article{Bigongiari:2011aa,
	author = {Bigongiari, A. and Raynaud, M. and Riconda, C. and H{\'e}ron, A. and Macchi, A.},
	doi = {10.1063/1.3646520},
	isbn = {1070-664X},
	journal = {Physics of Plasmas},
	journal1 = {Phys. Plasmas},
	month = {8/7/2025},
	number = {10},
	pages = {102701},
	title = {Efficient laser-overdense plasma coupling via surface plasma waves and steady magnetic field generation},
	url = {https://doi.org/10.1063/1.3646520},
	volume = {18},
	year = {2011},
	year1 = {2011/10/11}}

@article{Snavely:2000aa,
	author = {Snavely, R. A. and Key, M. H. and Hatchett, S. P. and Cowan, T. E. and Roth, M. and Phillips, T. W. and Stoyer, M. A. and Henry, E. A. and Sangster, T. C. and Singh, M. S. and Wilks, S. C. and MacKinnon, A. and Offenberger, A. and Pennington, D. M. and Yasuike, K. and Langdon, A. B. and Lasinski, B. F. and Johnson, J. and Perry, M. D. and Campbell, E. M.},
	date = {2000/10/02/},
	day = {02},
	doi = {10.1103/PhysRevLett.85.2945},
	id = {10.1103/PhysRevLett.85.2945},
	j1 = {PRL},
	journal = {Physical Review Letters},
	journal1 = {Phys. Rev. Lett.},
	month = {10},
	number = {14},
	pages = {2945--2948},
	publisher = {American Physical Society},
	title = {Intense High-Energy Proton Beams from Petawatt-Laser Irradiation of Solids},
	url = {https://link.aps.org/doi/10.1103/PhysRevLett.85.2945},
	volume = {85},
	year = {2000}}

@article{Hsu:2010aa,
	author = {Hsu, Jang-Yu},
	doi = {10.1063/1.3322854},
	isbn = {1070-664X},
	journal = {Physics of Plasmas},
	journal1 = {Phys. Plasmas},
	month = {8/7/2025},
	number = {3},
	pages = {032104},
	title = {Relativistic theory of mode conversion at plasma frequency},
	url = {https://doi.org/10.1063/1.3322854},
	volume = {17},
	year = {2010},
	year1 = {2010/03/05}}

@article{Baeva:2006aa,
	author = {Baeva, T. and Gordienko, S. and Pukhov, A.},
	date = {2006/10/12/},
	day = {12},
	doi = {10.1103/PhysRevE.74.046404},
	id = {10.1103/PhysRevE.74.046404},
	j1 = {PRE},
	journal = {Physical Review E},
	journal1 = {Phys. Rev. E},
	month = {10},
	number = {4},
	pages = {046404--},
	publisher = {American Physical Society},
	title = {Theory of high-order harmonic generation in relativistic laser interaction with overdense plasma},
	url = {https://link.aps.org/doi/10.1103/PhysRevE.74.046404},
	volume = {74},
	year = {2006}}

@article{Melrose:2006aa,
	author = {Melrose, D B and Weise, J I and McOrist, J},
	date = {2006/06/21},
	doi = {10.1088/0305-4470/39/27/011},
	isbn = {0305-4470},
	journal = {Journal of Physics A: Mathematical and General},
	number = {27},
	pages = {8727},
	title = {Relativistic quantum plasma dispersion functions},
	url = {https://dx.doi.org/10.1088/0305-4470/39/27/011},
	volume = {39},
	year = {2006}}

@article{Linde:1996aa,
	author = {von der Linde, D. and Rz{\`a}zewski, K.},
	date = {1996/11/01},
	doi = {10.1007/BF01828947},
	id = {von der Linde1996},
	isbn = {1432-0649},
	journal = {Applied Physics B},
	number = {5},
	pages = {499--506},
	title = {High-order optical harmonic generation from solid surfaces},
	url = {https://doi.org/10.1007/BF01828947},
	volume = {63},
	year = {1996}}

@article{Lichters:1996aa,
	author = {Lichters, R. and Meyer‐ter‐Vehn, J. and Pukhov, A.},
	doi = {10.1063/1.871619},
	isbn = {1070-664X},
	journal = {Physics of Plasmas},
	journal1 = {Phys. Plasmas},
	month = {8/7/2025},
	number = {9},
	pages = {3425--3437},
	title = {Short‐pulse laser harmonics from oscillating plasma surfaces driven at relativistic intensity},
	url = {https://doi.org/10.1063/1.871619},
	volume = {3},
	year = {1996},
	year1 = {1996/09/01}}

@article{Gibbon:1996aa,
	author = {Gibbon, Paul},
	date = {1996/01/01/},
	day = {01},
	doi = {10.1103/PhysRevLett.76.50},
	id = {10.1103/PhysRevLett.76.50},
	j1 = {PRL},
	journal = {Physical Review Letters},
	journal1 = {Phys. Rev. Lett.},
	month = {01},
	number = {1},
	pages = {50--53},
	publisher = {American Physical Society},
	title = {Harmonic Generation by Femtosecond Laser-Solid Interaction: A Coherent ``Water-Window'' Light Source?},
	url = {https://link.aps.org/doi/10.1103/PhysRevLett.76.50},
	volume = {76},
	year = {1996}}

@article{Bulanov:1994aa,
	author = {Bulanov, S. V. and Naumova, N. M. and Pegoraro, F.},
	doi = {10.1063/1.870766},
	isbn = {1070-664X},
	journal = {Physics of Plasmas},
	journal1 = {Phys. Plasmas},
	month = {8/7/2025},
	number = {3},
	pages = {745--757},
	title = {Interaction of an ultrashort, relativistically strong laser pulse with an overdense plasma},
	url = {https://doi.org/10.1063/1.870766},
	volume = {1},
	year = {1994},
	year1 = {1994/03/01}}

@article{Popov:1990aa,
	author = {Popov, E. and Tsonev, L.},
	date = {1990/05/01/},
	doi = {https://doi.org/10.1016/0039-6028(90)90037-9},
	isbn = {0039-6028},
	journal = {Surface Science},
	number = {1},
	pages = {290--294},
	title = {Total absorption of light by metallic gratings and energy flow distribution},
	url = {https://www.sciencedirect.com/science/article/pii/0039602890900379},
	volume = {230},
	year = {1990}}

@article{Carman:1981aa,
	author = {Carman, R. L. and Forslund, D. W. and Kindel, J. M.},
	date = {1981/01/05/},
	day = {05},
	doi = {10.1103/PhysRevLett.46.29},
	id = {10.1103/PhysRevLett.46.29},
	j1 = {PRL},
	journal = {Physical Review Letters},
	journal1 = {Phys. Rev. Lett.},
	month = {01},
	number = {1},
	pages = {29--32},
	publisher = {American Physical Society},
	title = {Visible Harmonic Emission as a Way of Measuring Profile Steepening},
	url = {https://link.aps.org/doi/10.1103/PhysRevLett.46.29},
	volume = {46},
	year = {1981}}

@article{Mourou2019aa,
	author = {Mourou, Gerard},
	doi = {10.1103/RevModPhys.91.030501},
	issue = {3},
	journal = {Rev. Mod. Phys.},
	month = {Jul},
	numpages = {21},
	pages = {030501},
	publisher = {American Physical Society},
	title = {Nobel Lecture: Extreme light physics and application},
	url = {https://link.aps.org/doi/10.1103/RevModPhys.91.030501},
	volume = {91},
	year = {2019}}

@article{Strickland:1985aa,
	author = {Strickland, Donna and Mourou, Gerard},
	date = {1985/12/01/},
	doi = {https://doi.org/10.1016/0030-4018(85)90120-8},
	isbn = {0030-4018},
	journal = {Optics Communications},
	number = {3},
	pages = {219--221},
	title = {Compression of amplified chirped optical pulses},
	url = {https://www.sciencedirect.com/science/article/pii/0030401885901208},
	volume = {56},
	year = {1985}}

@article{Umstadter:2003aa,
	author = {Donald Umstadter},
	date = {2003/04/02},
	doi = {10.1088/0022-3727/36/8/202},
	isbn = {0022-3727},
	journal = {Journal of Physics D: Applied Physics},
	number = {8},
	pages = {R151},
	title = {Relativistic laser--plasma interactions},
	url = {https://dx.doi.org/10.1088/0022-3727/36/8/202},
	volume = {36},
	year = {2003}}

@article{Vedenov:1965aa,
	author = {A A Vedenov},
	date = {1965/06/30},
	doi = {10.1070/PU1965v007n06ABEH003686},
	isbn = {0038-5670},
	journal = {Soviet Physics Uspekhi},
	number = {6},
	pages = {809},
	title = {SOLID STATE PLASMA},
	url = {https://dx.doi.org/10.1070/PU1965v007n06ABEH003686},
	volume = {7},
	year = {1965}}

@article{Otto:1968aa,
	author = {Otto, Andreas},
	date = {1968/08/01},
	doi = {10.1007/BF01391532},
	id = {Otto1968},
	isbn = {0939-7922},
	journal = {Zeitschrift f{\"u}r Physik A Hadrons and nuclei},
	number = {4},
	pages = {398--410},
	title = {Excitation of nonradiative surface plasma waves in silver by the method of frustrated total reflection},
	url = {https://doi.org/10.1007/BF01391532},
	volume = {216},
	year = {1968}}

@article{Ritchie:1957aa,
	author = {Ritchie, R. H.},
	date = {1957/06/01/},
	day = {01},
	doi = {10.1103/PhysRev.106.874},
	id = {10.1103/PhysRev.106.874},
	j1 = {PR},
	journal = {Physical Review},
	journal1 = {Phys. Rev.},
	month = {06},
	number = {5},
	pages = {874--881},
	publisher = {American Physical Society},
	title = {Plasma Losses by Fast Electrons in Thin Films},
	url = {https://link.aps.org/doi/10.1103/PhysRev.106.874},
	volume = {106},
	year = {1957}}

@article{Fano:1941aa,
	author = {Fano, U.},
	date = {1941/03/01},
	doi = {10.1364/JOSA.31.000213},
	j2 = {J. Opt. Soc. Am.},
	journal = {Journal of the Optical Society of America},
	journal1 = {Journal of the Optical Society of America},
	journal2 = {J. Opt. Soc. Am.},
	journal3 = {J. Opt. Soc. Am.},
	number = {3},
	pages = {213--222},
	publisher = {Optica Publishing Group},
	title = {The Theory of Anomalous Diffraction Gratings and of Quasi-Stationary Waves on Metallic Surfaces (Sommerfeld's Waves)},
	url = {https://opg.optica.org/abstract.cfm?URI=josa-31-3-213},
	volume = {31},
	year = {1941}}

@article{Wood:1902aa,
	author = {Wood, R. W.},
	date = {1902/09/01},
	doi = {10.1080/14786440209462857},
	isbn = {1941-5982},
	journal = {The London, Edinburgh, and Dublin Philosophical Magazine and Journal of Science},
	journal1 = {The London, Edinburgh, and Dublin Philosophical Magazine and Journal of Science},
	journal2 = {The London, Edinburgh, and Dublin Philosophical Magazine and Journal of Science},
	month = {09},
	number = {21},
	pages = {396--402},
	publisher = {Taylor \& Francis},
	title = {XLII. On a remarkable case of uneven distribution of light in a diffraction grating spectrum},
	type = {doi: 10.1080/14786440209462857},
	url = {https://doi.org/10.1080/14786440209462857},
	volume = {4},
	year = {1902},
	year1 = {1902}}

@article{Ceccotti:2013aa,
	author = {Ceccotti, T. and Floquet, V. and Sgattoni, A. and Bigongiari, A. and Klimo, O. and Raynaud, M. and Riconda, C. and Heron, A. and Baffigi, F. and Labate, L. and Gizzi, L. A. and Vassura, L. and Fuchs, J. and Passoni, M. and Kv{\v e}ton, M. and Novotny, F. and Possolt, M. and Prok{\r u}pek, J. and Pro{\v s}ka, J. and P{\v s}ikal, J. and {\v S}tolcov{\'a}, L. and Velyhan, A. and Bougeard, M. and D'Oliveira, P. and Tcherbakoff, O. and R{\'e}au, F. and Martin, P. and Macchi, A.},
	date = {2013/10/28/},
	day = {28},
	doi = {10.1103/PhysRevLett.111.185001},
	id = {10.1103/PhysRevLett.111.185001},
	j1 = {PRL},
	journal = {Physical Review Letters},
	journal1 = {Phys. Rev. Lett.},
	month = {10},
	number = {18},
	pages = {185001--},
	publisher = {American Physical Society},
	title = {Evidence of Resonant Surface-Wave Excitation in the Relativistic Regime through Measurements of Proton Acceleration from Grating Targets},
	url = {https://link.aps.org/doi/10.1103/PhysRevLett.111.185001},
	volume = {111},
	year = {2013}}

@article{Jana:2024aa,
	author = {Jana, Kamalesh and Lad, Amit D. and Zhang, Guo-Bo and Li, Bo-Yuan and Kumar, V. Rakesh and Shaikh, Moniruzzaman and Ved, Yash M. and Chen, Min and Kumar, G. Ravindra},
	doi = {10.1063/5.0168816},
	isbn = {1070-664X},
	journal = {Physics of Plasmas},
	journal1 = {Phys. Plasmas},
	month = {7/18/2025},
	number = {3},
	pages = {033105},
	title = {Collimated hot electron generation from sub-wavelength grating target irradiated by a femtosecond laser pulse of relativistic intensity},
	url = {https://doi.org/10.1063/5.0168816},
	volume = {31},
	year = {2024},
	year1 = {2024/03/07}}

@article{Cerchez:2013aa,
	author = {Cerchez, M. and Giesecke, A. L. and Peth, C. and Toncian, M. and Albertazzi, B. and Fuchs, J. and Willi, O. and Toncian, T.},
	date = {2013/02/07/},
	day = {07},
	doi = {10.1103/PhysRevLett.110.065003},
	id = {10.1103/PhysRevLett.110.065003},
	j1 = {PRL},
	journal = {Physical Review Letters},
	journal1 = {Phys. Rev. Lett.},
	month = {02},
	number = {6},
	pages = {065003--},
	publisher = {American Physical Society},
	title = {Generation of Laser-Driven Higher Harmonics from Grating Targets},
	url = {https://link.aps.org/doi/10.1103/PhysRevLett.110.065003},
	volume = {110},
	year = {2013}}

\end{document}